\documentclass[11pt]{amsart}
\usepackage{latexsym,amssymb,amsmath,amscd,amsthm}
\numberwithin{equation}{section}
\theoremstyle{remark}
\newtheorem{theorem}{{\bf THEOREM}}[section]

\newtheorem{corollary}{{\bf COROLLARY}}[section]
\newtheorem{example}{{\bf EXAMPLE}}[section]
\newtheorem{proposition}{{\bf PROPOSITION}}[section]
\newcommand{\bq}{\begin{equation}}
\newcommand{\bea}{\begin{array}}
\newcommand{\eea}{\end{array}}

\newcommand{\ga}{\alpha}
\newcommand{\gep}{\epsilon}
\newcommand{\gD}{\Delta}
\newcommand{\gl}{\lambda}
\newcommand{\gL}{\Lambda}
\newcommand{\gb}{\beta}

\newcommand{\mf}{\mathfrak}
\newcommand{\mc}{\mathcal}

\newcommand{\go}{\omega}
\newcommand{\gO}{\Omega}
\newcommand{\gG}{\Gamma}
\newcommand{\gt}{\theta}
\newcommand{\gs}{\sigma}

\newcommand{\gag}{\gamma}
\newcommand{\gd}{\delta}
\newcommand{\pp}{\partial}

\newcommand{\olra}{\overleftrightarrow}

\newcommand{\tl}{\tilde}
\newcommand{\na}{\nabla}
\newcommand{\gk}{\kappa}

\newcommand{\bl}{\blacklozenge}
\newcommand{\bs}{\blacksquare}

\newcommand{\bgs}{\bigstar}

\newcommand{\gS}{\Sigma}

\newcommand{{\DDD}}{D\!\!\!\!\!\!-}

\newcommand{\bx}{\Box}

\title{REMARKS ON GEOMETRY AND THE QUANTUM POTENTIAL}

\author{Robert Carroll\\University of Illinois, Urbana, IL 61801}

\date{February, 2007\thanks{email: rcarroll@math.uiuc.edu}}

\begin{document}

\bibliographystyle{plain}

\begin{abstract} 
We gather material from many sources about the quantum
potential and its geometric nature.  The presentation is primarily expository but 
some new observations relating $Q, V,$ and 
$\psi$ are indicated.
\end{abstract}

\maketitle

\tableofcontents


\section{INTRODUCTION}
\renewcommand{\theequation}{1.\arabic{equation}}
\setcounter{equation}{0}

In \cite{c1} we surveyed many feature of the so called quantum potential (QP)
(cf. also \cite{c8,c2,c3,c5,c7,c9,c17,c19}).   Some matters were treated more thoroughly than others and we want to discuss here certain geometrical aspects
in more detail, some connections to nonlinear Schr\"odinger type equations,
and various phase space approaches.  The latter two topics were not developed
in \cite{c1} and we will try to make amends here.  Some relations to electromagnetic (EM) theory will also be discussed.
To set the state we recall the
Schr\"odinger equation (SE) in 1-D of the form $({\bf 1A})\,\,-(\hbar^2/2m)\psi''+V\psi
=i\hbar\psi_t$ so that for $\psi=Rexp(iS/\hbar)$ one has
\bq\label{1.1}
S_t+\frac{1}{2m}S_x^2+V+Q=0;\,\,Q=-\frac{\hbar^2R''}{2mR};\,\,\pp_t(R^2)+\frac
{1}{m}(R^2S_x)_x=0
\end{equation}
Here Q is the quantum potential (QP) and one can argue that Bohmian mechanics is
simply classical symplectic mechanics using the Hamiltonian $({\bf 1B})\,\,H_q=
H_c+Q=(1/2m)S_x^2+V+Q$ from the Hamiltonian-Jacobi (HJ) equation (1.1) (cf.
here \cite{b2,b3,g1,h4}).  One can write $P=R^2=|\psi|^2$ (a probability density) with
$\rho=mP$ a mass density and obtain a hydrodynamical version of (1.1).  Note
in particular $({\bf 1C})\,\,Q=-(\hbar^2/2m)(\pp^2\sqrt{\rho}/\sqrt{\rho})$ and using
$p=S_x=m\dot{q}=mv$ one obtains
\bq\label{1.2}
mv_t+mvv_x+\pp V+\pp Q=0;\,\,\rho_t+(\rho\dot{q})_x=0
\end{equation}
leading to
\bq\label{1.3}
\pp_t(\rho v)+\pp(\rho v^2)+\frac{\rho}{m}\pp V+\frac{\rho}{m}\pp Q=0
\end{equation}
which has the flavor of an Euler equation (cf. \cite{c1,c7,d1}).  There is however a
missing pressure term from the hydrodynamical theory (cf. \cite{c1,l1,p5}) and looking at
(1.2) one could imagine a pressure term supplied in the form $({\bf 1D})\,\,\pp Q=
(1/R^2)\pp{\mf P}$ (where ${\mf P}$ denotes pressure).  This suggests a hydrodynamical interpretation for Q, namely, going to 3-D for example, $({\bf 1E})\,\,\na{\mf P}=R^2\na Q$ (cf. \cite{c7}).  This will all be discussed in detail below and we make first a few background remarks about the QP.
\\[3mm]\indent{\bf REMARK 1.1.}
In \cite{c8} we considered given a function $Q\in L^{\infty}(\gO)$ (for 
$\gO$ a bounded domain) and looked for $R\in H_0^1(\gO)$ satisfying
$Q=-(\hbar^2/2m)(\gD R/R)\equiv \gD R+(2m/\hbar^2)QR=0$.  We showed
that if $Q<0$ ($\gb=(2m/\hbar^2)$) then there is a unique solution and if
$0$ is not in the countable spectrum of $\gD R+\gb QR$ then $\gD R+\gb QR=0$ has a unique
solution for any $Q\in L^{\infty}$.  The corresponding HJ equation $({\bf 1F})\,\,
\pp_tS+(1/2m)(\na S)^2+Q+V=0$ and the continuity equation $({\bf 1G})\,\,
\pp_tR^2+(1/m)\na(R^2\na S)=0$ must then be solved to obtain some sort
of generalized quantum theory.  Here a priori V must be assumed unknown
and there are then two equations for two unknowns $S$ and $V$, namely
(in 1-D for simplicity)
\bq\label{1.4}
S_t+\frac{1}{2m}S_x^2+Q+V=0;\,\,\pp_tR^2+\frac{1}{m}(R^2S_x)_x=0
\end{equation}
the solution of which would yield a SE based on Q (see here Remark 5.1 for more detail in this regard).$\hfill\bs$  
\begin{example}
Now $(1/2m)p^2+
V=E$ (classical Hamiltonian - recall $p\sim S_x$) so we could perhaps treat
E as an unknown here and try to solve
\bq\label{1.5}
S_t+Q+E=0;\,\,\pp_tR^2+\frac{1}{m}(R^2S_x)_x=0
\end{equation}
Consider first $R^2S_x=-\int^xm\pp_tR^2dx+f(t)$ from which
\bq\label{1.6}
2RR_tS_x+R^2S_{xt}=-\int^xm\pp^2_tR^2dx+f'\Rightarrow (Q_x+E_x)R^2=
\end{equation}
$$=
-R^2S_{xt}=\frac{2R_t}{R}\left(-\int^xm\pp_tR^2dx+f\right)+\int^xm\pp^2_tR^2dx-f'$$
Hence 
\bq\label{1.7}
R^2S_x=-\int^xm\pp_tR^2dx+f(t);\,\,R^2E_x=-Q_xR^2+
\end{equation}
$$+\int^xm\pp^2_tR^2dx-f'+\frac{2R_t}{R}\left(-\int^xm\pp_tR^2dx+f\right)$$
giving $S_x$ and $E_x$ modulo an arbitrary differentiable function $f(t)$.
Note also
\bq\label{1.8}
R^2E_x=R^2V_x+\frac{R^2}{2m}(S_x^2)_x=R^2V_x+\frac{R^2}{m}S_xS_{xx}
\end{equation}
and $S_{xx}$ can be determined via $(1/m)(R^2S_x)_x=-\pp_tR^2$.  Hence
$R^2V_x$ can be determined from $R^2E_x$.  Note here 
\bq\label{1.88}
\frac{R^2}{m}S_xS_{xx}=-\pp_tR^2-\frac{2}{m}RR_xf+2RR_x\int^x\pp_tR^2dx
\end{equation}
(see Remark 5.1 for more details).
$\hfill\bs$
\end{example}
\indent
We mention also two examples from \cite{b2,c8}
\begin{example}
For a free particle in 1-D there are possibilities such as $\psi_1=Aexp[i[px-
(p^2t/2m))/\hbar]$ and $\psi_2=Aexp[-i(px+(p^2t/2m))/\hbar]$ in which case
$Q=0$ for both functions but for $\psi=(1/\sqrt{2})(\psi_1+\psi_2)$ there results
$Q=p^2/2m$ ($p\sim \hbar k$ here).  Hence $Q=0$ depends on the wave function
and cannot be said to represent a classical limit.  Further we note that
$S=\hbar kx-(\hbar^2k^2/2m)t$ in $\psi_1$ with $S_t=-\hbar^2k^2/2m\sim -E,\,\,
S_x=\hbar k$,
and $R=1\not\in H_0^1$.  For $\psi=(1/\sqrt{2})(\psi_1+\psi_2)$ on the other hand
\bq\label{1.9}
R=\sqrt{2}ACos(kx)\not\in H_0^1;\,\,\frac{R''}{R}=-k^2;\,\,Q=\frac{k^2\hbar^2}{2m};\,\,S=-\frac{k^2\hbar^2 t}{2m};
\end{equation}
$$S_t=-\frac{k^2\hbar^2}{2m}\sim -E;\,\,S_x=0$$
Thus the same SE can arise from different Q (which is generally obvious of course)
and S varies with Q.
$\hfill\bs$
\end{example}
\begin{example}
For $V=m\go^2x^2/2$ and a stationary SE one has solutions of the form 
$\psi_n(x)=c_nH_n(\xi x)exp(-\xi^2x^2/2)$ where $\xi=(m\go\hbar)^{1/2},\,\,
c_n=(\xi/\sqrt{\pi}2^nn!),$ and $H_n$ is a Hermite function.  One computes that
$Q=\hbar\go[n+(1/2)]-(1/2)m\go^2x^2$ and hence $\hbar\to 0$ does not imply
$Q\to 0$ and moreover $Q=0$ corresponds to $x=\pm\sqrt{(2\hbar/m\go)[n+(1/2)]}$ so not all systems in quantum mechanics have a classical limit.
This example corresponds to $\gO={\bf R}$ and $\psi_n\in H_0^1$ is satisfied.
$\hfill\bs$
\end{example}

\section{REMARKS ON WEYL GEOMETRY}
\renewcommand{\theequation}{2.\arabic{equation}}
\setcounter{equation}{0}

Now we recall how in various situations the QP is proportional to a Weyl-Ricci
curvature $R_w$ for example (cf. \cite{c1,c3,c6,s1}) and this can be interpreted in terms of a statistical geometry for example (cf. also \cite{a1,a5})
In general (see e.g. \cite{b2,c1,c8})
one knows that each wave function $\psi=Rexp(iS/\hbar)$ for a given SE produces a different QP as in (1.1) (which in higher dimensions has the form $-(\hbar^2/2m)
(\gD R/R)$).  Thus for $Q\sim R_w$ to make sense we have to think of a given R
or $R^2=P$ (or $\rho\sim mP$) as generating a (Weyl) geometry as in \cite{s1}.
(cf. also \cite{c1,c3,c6}).  
This is in accord with having a Weyl vector $({\bf 2A})\,\,
\phi_i\sim-\pp_i log(\hat{\rho})$ (where $\hat{\rho}=\rho/\sqrt{g}$ in \cite{s1} for
a Riemannian metric $g$).  Thus following \cite{s1} one assumes that the motion of
the particle is given by some random process $q^i(t,\go)$ in a manifold M 
($\go$ is the random process label) with a probability density $\rho(q,t)$ and satisfying
a deterministic equation $({\bf 2B})\,\,\dot{q}^i(t,\go)=(dq^i/dt)(t,\go)=v^i(q(t,\go),t)$ with
random initial conditions $q^i(t_0,\go)=q_0^i(\go)$.  The probability density will satisfy
$({\bf 2C})\,\,\pp_t\rho+\pp_i(\rho v^i)=0$ with initial data $\rho_0(q)$.  Let $L(q,\dot{q},t)$ be
some Lagrangian for the particle and define an equivalent Lagrangian via
\bq\label{2.1}
L^*(q,\dot{q},t)=L(q,\dot{q},t)-\pp_tS+q^i\pp_iS
\end{equation}
for some function S.  The velocity field $v^i(q,t)$ yielding a classical motion with probability
one can be found by minimizing the action functional
\bq\label{2.2}
I(t_0,t_1)=E\left[\int_{t_0}^{t_1}L^*(q(t,\go),\dot{q}(t,\go),t)dt\right]
\end{equation}
This leads to $({\bf 2D})\,\,\pp_tS+H(q,\na S,t)=0$  
and $p_i=(\pp L/\pp \dot{q}^i)=\pp_tS$ where $H\sim p_i\dot{q}^i-L$ with
$v^i(q,t)=(\pp H/\pp p_i)(q,\na S(q,t),t)$.  Now suppose that some geometric structure
is given on M via $ds^2=g_{ij}dq^idq^j$ so that a scalar curvature ${\mc R}(q,t)$ is meaningful and 
write the acutal Lagrangian as $({\bf 2E})\,\,L=L_C+\gag(\hbar^2/m){\mc R}(q,t)$
where $\gag$ will turn out to have the form $\gag=(1/8)[(n-2)/(n-1)]=1/16$ for $n=3$.
Assume that in a transplantation $q^i\to q^i+dq^i$ the length of a vector $\ell=
(g_{ik}A^iA^k)^{1/2}$ varies according to the law $({\bf 2F})\,\,\gd\ell=\ell\phi_kdq^k$
where the $\phi_k$ are covariant components of an arbitrary vector of M (this characterizes
a Weyl geometry).  One imagines that physics determines geometry so that the $\phi_k$ must be determined from some averaged least action principle yielding the motion of the particle;
in particular the minimum now in (2.2) is to be evaluated with respect to the class of all Weyl geometries with fixed metric tensor.  Since the only term containing the gauge vector
$\vec{\phi}=(\phi_k)$ is the curvature term one requires $E[{\mc R}(q(t,\go)t]=minimum$
($\gag>0$ for $n\geq 3$).  This minimization yields
\bq\label{2.3}
{\mc R}=\dot{{\mc R}}+(n-1)\left[(n-2)\phi_i\phi^i-2\left(\frac{1}{\sqrt{g}}\pp_i(\sqrt{g}\phi^i)
\right)\right]
\end{equation}
where $\phi^i=g^{ik}\phi_k$ and $\dot{{\mc R}}$ is the Riemannian curvature based on the metric.  Note here that a Weyl geometry is assumed as the proper background for the motion.
One shows that the quantity $\hat{\rho}(q,t)=\rho(q,t)/\sqrt{g}$ transforms as a 
scalar under coordinate changes and a covariant equation of the form $({\bf 2G})\,\,\pp_t\hat{\rho}
+(1/\sqrt{g})\pp_i(\sqrt{g}v^i\hat{\rho})=0$ ensues ($g_{ik}$ is assumed time independent).
Some calculation gives then a minimum when $({\bf 2H})\,\,\phi_i(q,t)=-[1/(n-2)]\pp_i
log(\hat{\rho})]$.  This shows that the transplantation properties of space are determined
by the presence of matter and in turn this change in geometry acts on the particle via a 
``quantum" force $f_i=\gag(\hbar^2/m)\pp_i{\mc R}$ depending on the gauge vector 
$\vec{\phi}$.  Putting this $\vec{\phi}$ in (2.3) yields
\bq\label{2.4}
R_w={\mc R}=\dot{{\mc R}}+\frac{1}{2\gag\sqrt{\hat{\rho}}}\left[\frac{1}{\sqrt{g}}\pp_i
(\sqrt{g}g^{ik}\pp_k\sqrt{\hat{\rho}}\right]
\end{equation}
along with a (HJ) equation
\bq\label{2.5}
\pp_tS+H_C(q,\na S,t)-\gag\left(\frac{\hbar^2}{m}\right){\mc R}=0
\end{equation}
and for certain Hamiltonians of the form $({\bf 2I})\,\,H_C=(1/2m)g^{ik}
(p_i-A_i)(p_k-A_k)+V$ with arbitrary fields $A_k$ and V it is shown that the function
$\psi=\sqrt{\hat{\rho}}exp[(i/\hbar)S(q,t)]$ satisfies a SE (omitting the $A_i$)
\bq\label{2.6}
i\hbar\pp_t\psi=-\frac{\hbar^2}{2m}\frac{1}{\sqrt{g}}\left[\pp_i\left(\sqrt{g}g^{ik}
\pp_k\right)\right]\psi+\left[V-\gag\left(\frac{\hbar^2}{m}\right)\dot{{\mc R}}\right]
\psi
\end{equation}
This Hamiltonian is characteristic of a particle in an EM field and all Hamiltonians arising
in nonrelativistic applications may be reduced to the above form with corresponding HJ
equation 
\bq\label{2.7}
\pp_tS=\frac{1}{2m}g^{ik}\pp_iS\pp_kS+V-\gag\frac{\hbar^2}{m}{\mc R}=0
\end{equation}
(note there are mistakes in the SE in \cite{s1} and in the improperly corrected form of 
\cite{c1}).  
\\[3mm]\indent
{\bf REMARK 2.1.}
Note that indices are lowered or raised via use of $g_{ij}$ or its inverse $g^{ij}$.
The most complete sources of notation for differential calculus on Riemannian
manifolds seem to be \cite{a6,w12}.  It is seen that $\hbar$ arises only via (2.5) and for
$\dot{{\mc R}}=0$ there is no $\hbar$ in the SE.  If ${\mc R}=0$ the quantum force is zero
and ``quantum mechanics" involves no $\hbar$; ${\mc R}=0$ (with $\dot{{\mc R}}=0$) 
involves (2.10) below giving $Q=0$.
$\hfill\bs$
\\[3mm]\indent
Now given (2.7), and comparing to (1.1) for example, we see that $({\bf 2J})\,\,Q\sim
-\gag(\hbar^2/m){\mc R}$ with ${\mc R}$ given by (2.4) and $\gag=1/16$ for $n=3$.  Thus
\bq\label{2.8}
Q\sim -\frac{\hbar^2}{16m}\left[\dot{{\mc R}}+\frac{8}{\sqrt{\hat{\rho} g}}\pp_i(\sqrt{g}g^{ik}
\pp_k\sqrt{\hat{\rho}})\right]
\end{equation}
and the SE (2.6) contains only $\dot{{\mc R}}$.  Further from ({\bf 2H}) we have for the Weyl
vector $\phi_i=-\pp_ilog(\hat{\rho})=-\pp_i\hat{\rho}/\hat{\rho}$ and there is an expression
for ${\mc R}$ in the form (2.3) leading to
\bq\label{2.9}
Q\sim-\frac{\hbar^2}{16m}\left[\dot{{\mc R}}+2\left\{\phi_i\phi^i-\frac{2}{\sqrt{g}}
\pp_i(\sqrt{g}\phi^i)\right\}\right]
\end{equation}
showing how Q depends directly on the Weyl vector.  When $\dot{{\mc R}}=0$ (flat space)
one sees that the SE is classical and $({\bf 2K})\,\,Q=-(\hbar^2/8m)[\phi_i\phi^i
-(1/\sqrt{g})\pp_i(\sqrt{g}\phi^i)]$.  Note that when $g=1$ (so $\dot{{\mc R}}=0$ automatically)
and $\hat{\rho}=\rho$ we have then
\bq\label{2.10}
\phi_k\phi^k-2\pp_k\phi^k\sim -\left(\frac{|\na\rho|^2}{\rho^2}-\frac{2\gD\rho}{\rho}\right)
=4\frac{\gD\sqrt{\rho}}{\sqrt{\rho}}
\end{equation}
which means $({\bf 2L})\,\,Q=-(\hbar^2/2m)(\gD\sqrt{\rho}/\sqrt{\rho})$ as in the desired ({\bf 1C}).
\\[3mm]\indent
{\bf REMARK 2.2}
Thus starting with a manifold M with metric $g_{ij}$ and random initial conditions
as indicated for a particle of mass $m$, the resulting classical statistical dynamics
based on a probability distribution P with $\rho=mP$ can be properly phrased in a Weyl geometry
in which the particle undergoes classical motion with probability one.  The 
assumed Weyl geometry as well as the particle motion is determined via $\hat{\rho}(\rho,g)$ which
says that given a different P there will be a different $\rho$ and $\hat{\rho}$
(since $g$ is fixed).  Hence writing $\psi=\sqrt{\hat{\rho}}exp(iS/\hbar)$ one expects
a different quantum potential and a different Weyl geometry.  The SE will however
remain unchanged and this may be a solution to the apparent problems illustrated
in Section 1 about different quantum potentials being attached to the same SE.
Another point of view could be that for $m$ fixed each $P$ (or equivalently $\rho$) determines a P-dependent motion via it Weyl geometry and each such motion can be
described by a P-dependent wave function.  The choice of $\hbar$ is arbitrary; here it
arises via $\psi$ and any $\hbar$ will do.  The identification with Planck's constant has
to come from other considerations.
$\hfill\bs$
\\[3mm]\indent
We go now to the second paper in \cite{s1} and sketch an interesting role of Weyl geometry
in the Klein-Gordon (KG) equation (cf. also \cite{c1,c19,c3,c6} for discussion of this approach).
The idea is to start from first principles, extended to gauge invariance relative to an arbitrary
choice of spacetime calibration.  Weyl geometry is not assumed but derived with the
particle motion from a single average action principle.  Thus assume a generic 4-D manifold
with torsion free connection $\gG^{\gl}_{\mu\nu}=\gG^{\gl}_{\nu\mu}$ and a metric tensor
$g$ with signature $(+,-,-,-)$; $\hbar=c=1$ is taken for convenience (although this 
loses important information in the equations).  The analysis will produce an integrable
Weyl geometry with weights $w(g_{\mu\nu})=1$ and $w(\gG^{\gl}_{\mu\nu})=0$
(cf. \cite{c1,i3} for Weyl geometry and Weyl-Dirac theory).
One takes random initial conditions on a spacelike 3-D hypersurface and produces both
particle motion and spacetime geometry via an average stationary action principle
$({\bf 2M})\,\,\gd\left[E\int_{\tau_1}^{\tau_2}L(x(\tau),\dot{x}(\tau))d\tau\right]=0$
where $\tau$ is an arbitrary parameter along the particle trajectory.  Given L positively
homogeneous of first degree in $\dot{x}^{\mu}=dx^{\mu}/d\tau$ and transforming as a scalar of
weight $w(L)=0$ as well as a gauge invariant probability measure it follows that the action
integral will be parameter invariant, coordinate invariant, and gauge invariant.  A suitable 
Lagrangian is $({\bf 2N})\,\,L(x,dx)=(m^2-(1/6){\mc R})^{1/2}ds+A_{\mu}dx^{\mu}$ where
$ds=(g_{\mu\nu}\dot{x}^{\mu}\dot{x}^{\nu})^{1/2}d\tau$ and $w(m)=-1/2$ ($m=$ rest mass
corresponds to a scalar Weyl field with no equation needed and the factor (1/6) in L is
for convenience later).  One writes $({\bf 2O})\,\,A_{\mu}=\bar{A}_{\mu}-\pp_{\mu}S$ where
$\bar{A}_{\mu}\sim$ EM 4-potential in Lorentz gauge and $w(S)=w(\bar{A}_{\mu})=0$.
\\[3mm]\indent
Omitting here the considerable details of calculation (which are given in \cite{s1} and
sketched in \cite{c1,c6}) one can work with a modified Lagrangian $({\bf 2P})\,\,\bar{L}(x,dx)=
(m^2-(1/6){\mc R})^{1/2}+\bar{A}_{\mu}dx^{\mu}$.  Variational methods lead to a 1-parameter
family of hypersurfaces $S(x)=constant$ satisfying the HJ equation
\bq\label{2.11}
g^{\mu\nu}(\pp_{\mu}S-\bar{A}_{\mu})(\pp_{\nu}S-\bar{A}_{\nu})=m^2-(1/6){\mc R}
\end{equation}
and a congruence of curves intersecting this family given via
\bq\label{2.12}
\frac{dx^{\mu}}{ds}=\frac{g^{\mu\nu}(\pp_{\nu}S-\bar{A}_{\nu})}{[g^{\rho\gs}(\pp_{\rho}-
\bar{A}_{\rho})(\pp_{\gs}S-\bar{A}_{\gs})]^{1/2}}
\end{equation}
The probability measure is determined by its probability current density $j^{\mu}$ where
$\pp_{\mu}j^{\mu}=0$ and $({\bf 2Q})\,\,j^{\mu}=\rho(\sqrt{-g}g^{\mu\nu}(\pp_{\nu}S-\bar{A}_{\nu})$.  Gauge invariance implies $w(j^{\mu})=0=w(S)$ and $w(\rho)=-1$ so $\rho$ is
the scalar probability density of the particle random motion.  To find the connection the
variational principle for ({\bf 2M}) is rephrased as
\bq\label{2.13}
\gd\left[\int_{\gO}d^4x[(m^2-(1/6){\mc R})(g_{\mu\nu}j^{\mu}j^{\nu})]^{1/2}+A_{\mu}j^{\mu}\right]=0
\end{equation}
Since the $\gG^{\gl}_{\mu\nu}$ arise only in ${\mc R}$ this reduces to $({\bf 2R})\,\,
\gd[\int_{\gO}\rho{\mc R}\sqrt{-g}d^4x]=0$ where (2.11) has been used.  This leads to
\bq\label{2.14}
\gG^{\gl}_{\mu\nu}=\left\{\begin{array}{c}
\gl\\
\mu\nu
\end{array}\right\}+\frac{1}{2}(\phi_{\mu}\gd^{\gl}_{\nu}+\phi_{\nu}\gd^{\gl}_{\mu}-g_{\mu\nu}
g^{\gl\rho}\phi_{\rho});\,\,\phi_{\mu}=\pp_{\mu}log(\rho)
\end{equation}
and shows that the connections are integrable Weyl connections with a 
gauge field $\phi_{\mu}$ (({\bf 2A}) suggests here perhaps $\phi_i=-(1/2)\pp_ilog(\rho)$).
The HJ equation (2.11) and $\pp_{\mu}j^{\mu}=0$ can be combined into a single equation
for $S(x)$, namely
\bq\label{2.15}
e^{iS}g^{\mu\nu}(iD_{\mu}-\bar{A}_{\mu})(iD_{\nu}-\bar{A}_{\nu})e^{-iS}-(m^2-(1/6){\mc R})=0
\end{equation}
with $D_{\mu}\rho=0$ where (cf. \cite{a1,c1})
\bq\label{2.16}
D_{\mu}T^{\ga}_{\gb}=\pp_{\mu}T^{\ga}_{\gb}+\gG^{\ga}_{\mu\gep}T^{\gep}_{\gb}-
\gG^{\gep}_{\mu\gb}T^{\ga}_{\gep}+w(T)\phi_{\mu}T^{\ga}_{\gb}
\end{equation}
($D_{\mu}$ is called the double-covariant Weyl derivative
and one notes that it is $\rho$ and not $m$, as in \cite{a1}, which behaves as a constant
under $D_{\mu}$).  Then to any solution $(\rho,S)$
of these equations corresponds a particular random motion for the particle.  One notes 
that (2.15)-(2.16) can be written in a familiar KG form
\bq\label{2.17}
\left(\frac{i}{\sqrt{-g}}\pp_{\mu}\sqrt{-g}-\bar{A}_{\mu}\right)g^{\mu\nu}(i\pp_{\nu}-\bar{A}_{\nu})
\psi-(m^2-(1/6)\dot{{\mc R}})\psi=0
\end{equation}
where $\psi=\sqrt{\rho}exp(-iS)$ and $\dot{{\mc R}}$ is the Riemannian scalar curvature.
We have also from \cite{s1}
\bq\label{2.18}
{\mc R}=\dot{{\mc R}}-3\left[\frac{1}{2}g^{\mu\nu}\phi_{\mu}\phi_{\nu}+\frac{1}{\sqrt{-g}}
\pp_{\mu}\sqrt{-g}g^{\mu\nu}\phi_{\nu}\right]=\dot{{\mc R}}+{\mc R}_w
\end{equation}
in keeping also with \cite{c6}.
\\[3mm]\indent
{\bf REMARK 2.3.}
Note here $g^{\mu\nu}\phi_{\nu}=\phi^{\mu}$ so (2.18) gives for the last term 
$({\bf 2S})\,\,
{\mc R}_w=-3[(1/2)\phi_{\mu}\phi^{\mu}+(1/\sqrt{-g})\pp_{\mu}(\sqrt{-g}\phi^{\mu})]$ whereas (2.3) 
suggests here $(\bgs)\,\,-3[2\phi_{\mu}\phi^{\mu}-(2/\sqrt{-g})\pp_{\mu}(\sqrt{-g}\phi^{\mu})]$
which is similar to paper 3 of \cite{s1} in having a minus sign in the middle; 
we remark that a change $\phi_{\mu}\to -2\phi_{\mu}$ would produce some
agreement and will stay with (2.18) or equivalently ({\bf 2S}) due to calculations
in Remark 2.5.
$\hfill\bs$
\\[3mm]\indent
{\bf REMARK 2.4.}
We add here a few standard formulas involving derivatives; thus
\bq\label{2.19}
\na_{\mu}\gl^{\nu}=\pp_{\mu}\gl^{\nu}+\gG^{\mu}_{\rho\nu}\gl^{\rho};\,\,\na_{\mu}\gl^{\mu}=
\pp_{\mu}\gl^{\mu}+\gG^{\mu}_{\rho\mu}\gl^{\rho}\,\,(divergence);
\end{equation}
$$\na_{\mu}\gl_{\nu}=\pp_{\mu}-\gG^{\rho}_{\nu\mu}\gl_{\rho};\,\,\gG^{\mu}_{\rho\mu}=
\pp_{\rho}log(\sqrt{g});\,\,\na_m\gl^m=\frac{1}{\sqrt{g}}\pp_m(\sqrt{g}\gl^m)$$
Also from (2.17)
\bq\label{2.20}
\bx\sim\frac{1}{\sqrt{-g}}\pp_{\mu}(\sqrt{-g}g^{\mu\nu}\pp_{\nu})=\na_{\mu}g^{\mu\nu}\pp_{\nu}=\na_{\mu}\na^{\mu}
\end{equation}
since $\na^{\mu}\sim\pp^{\mu}$ acting on functions (one could use $|g|$ instead of $\pm g$).
$\hfill\bs$
\\[3mm]\indent
{\bf REMARK 2.5.}
We see that for $\bar{A}_{\mu}=0$ the HJ equation (2.11) has the form
$({\bf 2T})\,\,\pp_{\mu}S\pp^{\mu}S=m^2-(1/6){\mc R}$ and mention that it is shown in 
\cite{c6} that the $1/6$ factor is essential if one wants a linear KG equation.
We want now to identify ${\mc R}_w$ with a multiple of Q which should have a form
like $({\bf 2U})\,\,Q\propto(1/\sqrt{\rho}) \na^{\mu}\na_{\mu}(\sqrt{\rho})$.  A crude calculation suggests
\bq\label{2.21}
\pp_{\mu}\pp^{\mu}\sqrt{\rho}=\pp_{\mu}\left[\frac{1}{2}\rho^{-1/2}\pp^{\mu}\rho\right]=\frac{1}{2}
\left[-\frac{1}{2}\rho^{-3/2}\pp_{\mu}\rho\pp^{\mu}\rho+\rho^{-1/2}\pp_{\mu}\pp^{\mu}\rho\right]
\Rightarrow\end{equation}
$$\Rightarrow \frac{\pp_{\mu}\pp^{\mu}\sqrt{\rho}}{\sqrt{\rho}}=\frac{1}{2}\left[
-\frac{1}{2}\frac{\pp_{\mu}\rho\pp^{\mu}\rho}{\rho^2}+
\frac{\pp_{\mu}\pp^{\mu}\rho}{\rho}\right]$$
and it is easy to check (cf. \cite{o3}) that $\na_m(fg^m)=(\na_mf)g^m + f(\na_mg^m)$.
Hence $\na_m\na^m\sqrt{\rho}$ can be written out as in (2.21) to get
\bq\label{2.22}
\frac{\bx(\sqrt{\rho})}{\sqrt{\rho}}=\frac{1}{2}\left[-\frac{1}{2}\frac{\na_{\mu}\rho\na^{\mu}
\rho}{\rho^2}+\frac{\bx(\rho)}{\rho}\right]
\end{equation}
and hence from (2.18)
\bq\label{2.23}
{\mc R}_w=-3\left[\frac{1}{2}\frac{\na_{\mu}\rho\na^{\mu}\rho}{\rho^2}+\na_{\mu}
\left(\frac{\na^{\mu}\rho}{\rho}\right)\right]=-3\left[\frac{1}{2}\frac{\na_{\mu}\rho
\na^{\mu}\rho}{\rho^2}+\right.
\end{equation}
$$\left.+\frac{\na_{\mu}\na^{\mu}\rho}{\rho}-\frac{\na_{\mu}\rho\na^{\mu}\rho}{\rho^2}
\right]=-6\frac{\bx(\sqrt{\rho})}{\sqrt{\rho}}$$
The formula for Q is then $({\bf 2V})\,\,Q=-[\bx(\sqrt{\rho})/\sqrt{\rho}]
=(1/6){\mc R}_w$.  We remark that in various contexts formulas for Q arise here with
multipliers $1/m^2,\,\,\hbar^2/2m,$ etc. (cf. \cite{c1} and remarks below).$\hfill\bs$

\section{EMERGENCE OF Q IN GEOMETRY}
\renewcommand{\theequation}{3.\arabic{equation}}
\setcounter{equation}{0}

In \cite{c1} we have indicated a number of contexts where Q arises in geometrical
situations involving KG type equations and we review this here (cf. also \cite{c19}).
We list a number of
occasions (while omitting others).
\begin{enumerate}
\item
We omit any details for the Bertoldi-Faraggi-Matone (BFM) approach (see \cite{b1,c1,f9,f2,f6})
since it involves a whole philosophy (of considerable importance).  Thus for
$\eta^{\mu\nu}=diag(-1,1,1,1)$ and $q=(ct,q_1,q_2,q_3)$ one has
\bq\label{3.1}
\frac{1}{2m}\eta^{\mu\nu}\pp_{\mu}S^{cl}\pp_{\nu}S^{cl}+{\mf W}'_{rel}=0;
\end{equation}
with ${\mf W}'_{rel}=\frac{1}{2mc^2}[m^2c^4-V^2(q)-2cV(q)\pp_0S^{cl}]$ where V is some potential which we could take to be zero.
The quantum version attaches Q to (3.1) to get ($S^{cl}\to S$)
\bq\label{3.2}
\frac{1}{2m}(\pp S)^2+{\mf W}_{rel}+Q=0;\,\,{\mf W}_{rel}=\frac{1}{2mc^2}
[m^2c^4-V^2-2cV\pp_0S]
\end{equation}
This involves then
\bq\label{3.3}
{\mf W}_{rel}=\left(\frac{\hbar^2}{2m}\right)\frac{\bx (Re^{iS/\hbar})}{Re^{iS/\hbar}};\,\,
Q=-\frac{\hbar^2}{2m}\frac{\bx R}{R};\,\,\pp\cdot(R^2\pp S)=0
\end{equation}
where one uses $\pp\sim\na$ when $g_{\mu\nu}=\eta_{\mu\nu}$.
\item
One can derive the SE, the KG equation, and the Dirac equation using methods of
scale relativity (cf. \cite{c1,c19,c20,c21,n4,n8,p10}); here  e.g. quantum
paths are considered to be continuous nondifferentiable curves with left and right derivatives at any
point.  Using a ``diffusion" coefficient $D=\hbar/2m$ as in the Nelson theory
(cf. \cite{c1,c19,n6}) one defines ``average" velocities $V=(1/2)[d_{+}x(t)+
d_{-}x(t)]$ and $U=(1/2)[d_{+}x(t)-d_{-}x(t)]$.  Then e.g. there is a SE $i\hbar\psi_t
=-(\hbar^2/2m)\gD\psi+{\mf U}\psi$ with quantum potential $Q=-(m/2)U^2-
(\hbar/2)\pp U$ where $U=(\hbar/m)(\pp\sqrt{\rho}/\sqrt{\rho})$.  The ideas 
should be extendible to a KG equation where $Q\sim (\hbar^2/m^2c^2)
(\bx_g|\psi|/\|psi|)$ (see Section 5).
\item
One can construct directly a KG theory following \cite{n1} in the form $(\pp_0^2-\na^2+m^2)\phi=0$ where $\eta_{\mu\nu}=(1,-1,-1,-1)$.  If $\psi=\phi^{+}$ with $\psi^*=\phi^{-}$
correspond to positive and negative frequency parts of $\phi=\phi^{+}+\phi^{-}$ the
particle current is $j_{\mu}=i\psi^*\olra{\pp}_{\mu}\psi$ and $N=\int d^3xj_0$ is the 
particle number.  Trajectories have the form $d{\bf x}/dt={\bf j}(t,{\bf x})/j_0(t,{\bf x})$ for 
$t=x_0$ and for $c=\hbar=1$ one arrives at
\bq\label{3.4}
\pp^{\mu}(R^2\pp_{\mu}S)=0;\,\,\frac{(\pp^{\mu}S)(\pp_{\mu}S)}{2m}-\frac{m}{2}+Q=0;
\,\,Q=-\frac{1}{2m}\frac{\pp^{\mu}\pp_{\mu}R}{R}
\end{equation}
\item
A covariant field theoretic version is also given in \cite{n1} using deDonder-Weyl theory
(cf. also \cite{c1,c19}).  One works with a real scalar field $\phi(x)$ and defines
$({\bf 3A})\,\,{\mf A}=\int d^4x{\mf L};\,\,{\mf L}=(1/2)(\pp^{\mu}\phi)(\pp_{\mu}\phi)
-V(\phi)$ with $({\bf 3B})\,\,\pi^{\mu}=\pp{\mf L}/
\pp(\pp_{\mu}\phi)=\pp^{\mu}\phi,\,\,
\pp_{\mu}\phi=\pp{\mf H}/\pp\pi^{\mu}$, and $\pp_{\mu}\pi^{\mu}=-\pp{\mf H}/\pp\phi$.
One takes a preferred foliation of spacetime with $R^{\mu}$ normal to the leaf $\gS$ and writes ${\mf R}([\phi],\gS)=\int_{\gS}d\gS_{\mu}R^{\mu}$ with ${\mf S}([\phi],\gS)=\int_{\gS}d\gS_{\mu}S^{\mu}$ and $\Psi={\mf R}exp(i{\mf S}/\hbar)$.  A covariant
version of Bohmian mechanics ensues with 
\bq\label{3.5}
\frac{1}{2}\frac{dS_{\mu}}{d\phi}\frac{dS^{\mu}}{d\phi}+V+Q+\pp_{\mu}S^{\mu}=0;\,\,
\frac{dR^{\mu}}{d\phi}\frac{dS^{\mu}}{d\phi}+J+\pp_{\mu}R^{\mu}=0
\end{equation}
\bq\label{3.6}
Q=-\frac{\hbar^2}{2{\mf R}}\frac{\gd^2{\mf R}}{\gd_{\gS}\phi^2};\,\,J=\frac{{\mf R}}{2}
\frac{\gd^2{\mf S}}{\gd_{\gS}\phi^2}
\end{equation}
The nature of this approach as a covariant version of the Bohmian hidden variable
theory is spelled out in the last paper of \cite{n1}.  This is a significant extension of 
earlier classical field theoretic approaches and another lovely extension is described 
by Nikolic in \cite{n2} involving a covariant many fingered time Bohmian interpretation
of quantum field theory (QFT).
\end{enumerate}
\indent
We preface the next set of examples with a discussion of a formula ${\mf M}^2=m^2exp({\mf Q}_{rel})$ used in \cite{s2} in
an important manner and produced also in \cite{n9}.  This formula differs from the result
${\mf M}=mexp({\mf Q}_{rel})$ of \cite{s13} (which was abandoned in \cite{s2}) and in order to
clarify this we write out in more detail the approach of \cite{n9}.  Thus one is dealing with
a Bohmian theory and for a Klein-Gordon (KG) equation a wave function $\psi=Rexp(iS/\hbar)$
this leads to 
\bq\label{8.1}
\pp_{\mu}(R^2\pp^{\mu}S)=0;\,\,\pp_{\mu}S\pp^{\mu}S={\mf M}^2c^2\,\,(\sim m^2c^2(1+{\mf
Q}_{rel}))
\end{equation}
where ${\mf Q}_{rel}=(\hbar^2/m^2c^2)(\pp_{\mu}\pp^{\mu}R/R)$ and (temporarily  now)
$\pp_{\mu}\pp^{\mu}\sim\bx=(1/c^2)\pp_t^2-\gD$ where $\eta_{\ga\gb}=diag(1,-1,-1,-1)$.  Now
${\mf M}=1+{\mf Q}_{rel}$ is only an approximation (leading e.g. to tachyon problems) and a
better formula for ${\mf M}$ can be found as follows.  Thus one knows 
$({\bf 3C})\,\,
(dx^{\mu}(\tau)/d\tau)=(1/m)\pp^{\mu}S$ and differentiating gives
\bq\label{8.2}
\pp_{\tau}\pp^{\mu}S=\pp_{\nu}\pp^{\mu}S\frac{dx^{\mu}}{d\tau}=
\pp_{\nu}{\mf M}\frac{dx^{\nu}}{d\tau}\frac{dx^{\mu}}{d\tau}+{\mf M}\frac{d^2x^{\mu}}{d\tau^2}
\end{equation}
But via the formula (valid for $g_{ab}=\eta_{ab}$ constant
\bq\label{8.22}
\pp_b(\pp_aS\pp^aS)=(\pp_b\pp_aS)(\pp^aS)+(\pp_aS)(\pp_b\pp^aS);
\end{equation}
$$\pp_a\pp_b\pp^aS=\pp_aS\eta^{ac}\pp_c\pp_bS=\pp^cS\pp_c\pp_bS$$
one has
$({\bf 3D})\,\,\pp_{\nu}(\pp_{\mu}S\pp^{\mu}S)=2(\pp^{\mu}S)(\pp_{\mu}\pp_{\nu}S)$
and therefore
\bq\label{8.3}
\pp_{\nu}(\pp_{\mu}S\pp^{\mu}S)=\pp_{\nu}({\mf M}^2c^2)=2{\mf M}\pp_{\nu}{\mf M}c^2=2
(\pp^{\nu}S)(\pp_{\mu}\pp_{\nu}S)=
\end{equation}
$$=2{\mf M}\frac{dx^{\mu}}{d\tau}(\pp_{\mu}\pp_{\nu}S)$$
Hence $({\bf 3E})\,\,\pp_{\nu}{\mf M}c^2=(\pp_{\mu}\pp_{\nu}S)(dx^{\mu}/d\tau)$ which implies
\bq\label{8.4}
\eta^{\ga\nu}c^2\pp_{\nu}{\mf M}=\eta^{\ga\nu}\pp_{\mu}\pp_{\nu}S(dx^{\mu}/d\tau)=
\pp_{\mu}\pp^{\ga}S(dx^{\mu}/d\tau)
\end{equation}
Consequently \eqref{8.2} becomes 
\bq\label{8.5}
\eta^{\ga\nu}c^2\pp_{\nu}{\mf M}=\pp_{\nu}{\mf M}\frac{dx^{\nu}}{d\tau}\frac{dx^{\ga}}{d\tau}
+{\mf M}\frac{d^2x^{\ga}}{d\tau^2}\equiv
\end{equation}
$$\equiv
{\mf M}\frac{d^2x^{\ga}}{d\tau^2}=\left(c^2\eta^{\ga\nu}-
\frac{dx^{\nu}}{d\tau}\frac{dx^{\ga}}{d\tau}\right)\pp_{\nu}{\mf M}$$
and this is equation (9) of \cite{n9}.  
For $|\dot{x}^{\ga}|<<c$ one obtains then
${\mf M}\ddot{x}^{\ga}\sim c^2\pp^{\ga}{\mf M}\sim-c^2\pp_{\ga}{\mf M}$ and comparing with the
nonrelativistic equation $m\ddot{x}^{\ga}=-\pp_{\ga}Q_{cl}$ implies ${\mf M}\sim mexp
({\mf Q}_{cl}/mc^2)$ and suggests that ${\mf M}\sim mexp({\mf Q}_{rel}/2)$ (recall ${\mf
Q}_{cl}=-(\hbar^2/2m)(\na^2|\psi|/|\psi|))$.
\\[3mm]\indent
Now one observes that the quantum effects will affect the geometry and in
fact are equivalent to a change of spacetime metric
\bq\label{8.6}
g_{\mu\nu}\to\tl{g}_{\mu\nu}=({\mf M}^2/m^2)g_{\mu\nu}
\end{equation}
(conformal transformation).  The QHJE becomes 
$\tl{g}^{\mu\nu}\tl{\na}_{\mu}S\tl{\na}_{\nu}S=m^2c^2$ where $\tl{\na}_{\mu}$
represents covariant differentiation with respect to the metric
$\tl{g}_{\mu\nu}$ and the continuity equation is then 
$\tl{g}_{\mu\nu}\tl{\na}_{\mu}(\rho\tl{\na}_{\nu}S)=0$.
The important conclusion here is that the presence of the quantum potential is
equivalent to a curved spacetime with its metric given by (3.13).  This is a
geometrization of the quantum aspects of matter and it seems that there is a dual
aspect to the role of geometry in physics.  The spacetime geometry sometimes looks
like ``gravity" and sometimes reveals quantum behavior.  The curvature due to the
quantum potential may have a large influence on the classical contribution to the
curvature of spacetime.  The particle trajectory can now be derived from the
guidance relation via differentiation as in ({\bf 3C}) again, leading to the Newton
equations of motion
\bq\label{8.7}
{\mf M}\frac{d^2x^{\mu}}{d\tau^2}+{\mf M}\gG^{\mu}_{\nu\gk}u^{\nu}u^{\gk}=
(c^2g^{\mu\nu}-u^{\mu}u^{\nu})\na_{\nu}{\mf M}
\end{equation}
Using the conformal transformation above \eqref{8.7} reduces to the standard
geodesic equation.
\\[3mm]\indent
We extract now from \cite{c1,s2,s3,s4,s13,s14,s15,s16,s17,s18} with emphasis
on the survey article \cite{s2}.  This may seem overly repetitious but the material
seems worthy of further emphasis.
Thus a general ``canonical" relativistic system consisting of gravity and classical matter (no quantum effects) is determined by the action
\bq\label{8.8}
{\mc A}=\frac{1}{2\gk}\int d^4x\sqrt{-g}{\mc R}+\int
d^4x\sqrt{-g}\frac{\hbar^2}{2m}\left(\frac{\rho}{\hbar^2}{\mc D}_{\mu}S{\mc
D}^{\mu}S-\frac{m^2}{\hbar^2}\rho\right)
\end{equation}
where $\gk=8\pi G$ and $c=1$ for convenience and ${\mc D}_{\mu}$ is the covariant derivative based
on  
$g_{\mu\nu}$ (${\mc D}_{\mu}\sim\na_{\mu}$).  It was seen above that via
deBroglie the introduction of a quantum potential is equivalent to introducing a
conformal factor $\gO^2={\mf M}^2/m^2$ in the metric.  Hence in order to introduce
quantum effects of matter into the action \eqref{8.8} one uses this conformal
transformation to get ($1+Q\sim exp(Q)$ and $Q\sim(\hbar^2/c^2m^2)(\bx(\sqrt{\rho})
/\sqrt{\rho})$ with $c=1$ here)
\bq\label{8.9}
{\mf A}=\frac{1}{2\gk}\int d^4x\sqrt{-\bar{g}}(\bar{{\mc R}}\gO^2-6\bar{\na}_{\mu}
\gO\bar{\na}^{\mu}\gO)+
\end{equation}
$$+\int
d^4x\sqrt{-\bar{g}}\left(\frac{\rho}{m}\gO^2\bar{\na}_{\mu}S\bar{\na}^{\mu}S-
m\rho\gO^4\right)+$$
$$+\int
d^4x\sqrt{-\bar{g}}\gl\left[\gO^2-\left(1+\frac{\hbar^2}{m^2}
\frac{\stackrel{-}{\bx}\sqrt{\rho}}{\sqrt{\rho}}\right)\right]$$
where a bar over any quantity means that it corresponds to the nonquantum regime.
Here only the first two terms of the expansion of ${\mf M}^2=m^2exp({\mf Q})$ have been used,
namely ${\mf M}^2\sim m^2(1+{\mf Q})$.  $\gl$ is a Lagrange multiplier introduced to identify the
conformal factor with its Bohmian value.  One uses here $\bar{g}_{\mu\nu}$ to raise or lower
indices and to evaluate the covariant derivatives; the physical metric (containing the quantum
effects of matter) is
$g_{\mu\nu}=\gO^2\bar{g}_{\mu\nu}$.  By variation of the action with respect to
$\bar{g}_{\mu\nu},\,\gO,\,\rho,\,S,$ and $\gl$ one arrives at 
quantum equations of motion, including quantum Einstein equations (cf. \cite{c1,s2}).
There is a generalized equivalence principle.  The gravitational effects determine
the causal structure of spacetime as long as quantum effects give its conformal structure.
This does not mean that quantum effects have nothing to do with the causal structure; they
can act on the causal structure through back reaction terms appearing in the metric field
equations.  The conformal factor of the metric is a function of the quantum potential and
the mass of a relativistic particle is a field produced by quantum corrections to the
classical mass.  One has shown that the presence of the quantum potential is equivalent to a
conformal mapping of the metric.  Thus in different conformally related frames one ``feels"
different quantum masses and different curvatures.  In particular there are two frames with
one containing the quantum mass field and the classical metric while the other contains
the classical mass and the quantum metric.  In general frames both the spacetime metric and
the mass field have quantum properties so one can state that different conformal frames are
identical pictures of the gravitational and quantum phenomena.  One ``feels" different quantum
forces in different conformal frames.  The question then arises of whether the
geometrization of quantum effects implies conformal invariance just as gravitational effects
imply general coordinate invariance.  One sees here that Weyl geometry provides 
additional degrees of freedom which can be identified with quantum effects and seems
to create a unified geometric framework for understanding both gravitational and quantum
forces.  Some features here are: (i) Quantum effects appear independent of any preferred
length scale.  (ii) The quantum mass of a particle is a field.  (iii) The gravitational
constant is also a field depending on the matter distribution via the quantum potential.
(iv)  A local variation of matter field distribution changes the
quantum potential acting on the geometry and alters it globally; the nonlocal character is
forced by the quantum potential (cf. \cite{c1,s2,s17}).  
\\[3mm]\indent
Next (still following \cite{s2}) one goes to Weyl geometry based on the Weyl-Dirac action
\bq\label{8.10}
{\mf A}=\int d^4x\sqrt{-g}(F_{\mu\nu}F^{\mu\nu}-\gb^2\,\,{}^W{\mc
R}+(\gs+6)\gb_{;\mu}\gb^{;\mu}+ {\mf L}_{matter})
\end{equation}
Here $F_{\mu\nu}$ is the curl of the Weyl 4-vector $\phi_{\mu}$, $\gs$ is an arbitrary
constant and $\gb$ is a scalar field of weight $-1$.  The symbol
``;" represents a covariant derivative under general coordinate and conformal transformations
(Weyl covariant derivative) defined as $X_{;\mu}={}^W\na_{\mu}X-{\mc
N}\phi_{\mu}X$ where 
${\mc N}$ is the Weyl weight of X.  The equations of motion are then
given in \cite{c1,s2}.
There is then agreement with the Bohmian theory provided one identifies
\bq\label{8.11}
\gb\sim{\mf M};\,\,\frac{8\pi{\mf T}}{{\mc R}}\sim
m^2;\,\,\frac{1}{\gs\phi_{\ga}\phi^{\ga}-({\mc R}/6)}\sim \ga=\frac{\hbar^2}{m^2c^2}
\end{equation}
Thus $\gb$ is the Bohmian quantum mass field and the coupling constant $\ga$ (which depends
on $\hbar$) is also a field, related to geometrical properties of spacetime.  One notes that
the quantum effects and the length scale of the spacetime are related.  To see this suppose
one is in a gauge in which the Dirac field is constant; apply a gauge transformation to
change this to a general spacetime dependent function, i.e.
\bq\label{8.12}
\gb=\gb_0\to\gb(x)=
\gb_0exp(-\Xi(x));\,\,\phi_{\mu}\to\phi_{\mu}+\pp_{\mu}\Xi
\end{equation}
Thus the gauge in which the
quantum mass is constant (and the quantum force is zero) and the gauge in which the quantum
mass is spacetime dependent are related to one another via a scale change.  In particular
$\phi_{\mu}$ in the two gauges differ by $-\na_{\mu}(\gb/\gb_0)$ and since $\phi_{\mu}$ is a
part of Weyl geometry and the Dirac field represents the quantum mass one concludes that the
quantum effects are geometrized which shows that $\phi_{\mu}$ is not
independent of $\gb$ so the Weyl vector is determined by the quantum mass and thus the
geometrical aspects of the manifold are related to quantum effects).

\subsection{QUANTUM POTENTIAL AS A DYNAMICAL FIELD}

In \cite{s2,s17} (cf. also \cite{c1}) one can write down a scalar tensor theory where the
conformal factor and the quantum potential are both dynamical fields but first we deal
with (3.16). 
For the relativistic situation one will have e.g. ${\mf Q}=(\hbar^2/m^2c^2)
(\bx_g|\psi|/|\psi|)$ where $\bx_g|\psi|\sim \na_{\ga}\na^{\ga}|\psi|=g^{\ga\gb}\na_{\gb}\na_{\ga}
|\psi|$ and the HJ equation is $\na_{\mu}S\na^{\mu}S={\mf M}^2c^2$ where ${\mf M}^2=
m^2exp({\mf Q})$.  Equivalently $\tl{g}^{\mu\nu}\tl{\na}_{\mu}S\tl{\na}_{\nu}S=m^2c^2$ where 
$g_{\mu\nu}=({\mf M}/m)^2\tl{g}_{\mu\nu}$ and $\tl{\na}_{\mu}$ is the covariant derivative with
respect to $\tl{g}_{\mu\nu}$.  The corresponding geodesic equation is given via (3.14).  We
write $\gO^2=({\mf M}/m)^2$ and this leads to (3.16) based on the fundamental action (3.15).
Recall here $exp({\mf Q})\sim m^2(1+{\mf Q})$ has been used for ${\mf M}$ in the last term in
(3.16).  We recall also the fundamental equations determined by varying the action (3.16) with
respect to $\bar{g}_{\mu\nu},\,\gO,\,\rho,\,S,$ and $\gl$ are (cf. \cite{c1,s2})
\begin{enumerate}
\item
The equation of motion for $\gO$
\bq\label{9.1}
\bar{{\mc
R}}\gO+6\stackrel{-}{\bx}\gO+\frac{2\gk}{m}\rho\gO(\bar{\na}_{\mu}S\bar{\na}^{\mu}S-
2m^2\gO^2)+2\gk\gl\gO=0
\end{equation}
\item
The continuity equation for particles 
$\bar{\na}_{\mu}(\rho\gO^2\bar{\na}^{\mu}S)=0$
\item
The equations of motion for particles 
\bq\label{9.2}
(\bar{\na}_{\mu}S\bar{\na}^{\mu}S-m^2\gO^2)\gO^2\sqrt{\rho}+\frac{\hbar^2}{2m}
\left[\stackrel{-}{\bx}\left(\frac{\gl}{\sqrt{\rho}}\right)-\gl\frac{\stackrel{-}{\bx}\sqrt{\rho}}{\rho}\right]
=0
\end{equation}
\item
The modified Einstein equations for $\bar{g}_{\mu\nu}$
\bq\label{9.3}
\gO^2\left[\bar{{\mc R}}_{\mu\nu}-\frac{1}{2}\bar{g}_{\mu\nu}\bar{{\mc R}}\right]
-[\bar{g}_{\mu\nu}\stackrel{-}{\bx}-\bar{\na}_{\mu}\bar{\na}_{\nu}]\gO^2-6\bar{\na}_{\mu}
\gO\bar{\na}_{\nu}\gO+3\bar{g}_{\mu\nu}\bar{\na}_{\ga}\gO\bar{\na}^{\ga}\gO+
\end{equation}
$$+\frac{2\gk}{m}\rho\gO^2\bar{\na}_{\mu}S\bar{\na}_{\nu}S-\frac{\gk}{m}
\rho\gO^2\bar{g}_{\mu\nu}\bar{\na}_{\ga}S\bar{\na}^{\ga}S+\gk
m\rho\gO^4\bar{g}_{\mu\nu}+$$
$$+\frac{\gk\hbar^2}{m^2}\left[\bar{\na}_{\mu}\sqrt{\rho}\bar{\na}_{\nu}\left(\frac
{\gl}{\sqrt{\rho}}\right)+\bar{\na}_{\nu}\sqrt{\rho}\bar{\na}_{\mu}\left(\frac{\gl}{\sqrt{\rho}}
\right)\right]-\frac{\gk\hbar^2}{m^2}\bar{g}_{\mu\nu}\bar{\na}_{\ga}\left[
\gl\frac{\bar{\na}^{\ga}\sqrt{\rho}}{\sqrt{\rho}}\right]=0$$
\item
The constraint equation
$\gO^2=1+(\hbar^2/m^2)[(\stackrel{-}{\bx}\sqrt{\rho})/\sqrt{\rho}]$
\end{enumerate}
\indent
Thus the back reaction effects of the quantum factor on the background metric are
contained in these highly coupled equations.  A simpler form of
\eqref{9.1} can be obtained by taking the trace of \eqref{9.2} and using \eqref{9.1}
which produces 
$\gl=(\hbar^2/m^2)\bar{\na}_{\mu}[\gl(\bar{\na}^{\mu}\sqrt{\rho})/
\sqrt{\rho}]$.  A solution of this via perturbation methods using the small
parameter $\ga=\hbar^2/m^2$ yields the trivial solution $\gl=0$ so the above
equations reduce to
\bq\label{9.4}
\bar{\na}_{\mu}(\rho\gO^2\bar{\na}^{\mu}S)=0;\,\,\bar{\na}_{\mu}S\bar{\na}^{\mu}
S=m^2\gO^2;\,\,{\mf G}_{\mu\nu}=-\gk{\mf T}^{(m)}_{\mu\nu}-\gk{\mf
T}^{(\gO)}_{\mu\nu}
\end{equation}
where ${\mf T}_{\mu\nu}^{(m)}$ is the matter energy-momentum (EM) tensor and 
\bq\label{9.5}
\gk{\mf T}_{\mu\nu}^{(\gO)}=\frac{[g_{\mu\nu}\bx-\na_{\mu}\na_{\nu}]\gO^2}{\gO^2}
+6\frac{\na_{\mu}\gO\na_{\nu}\gO}{\go^2}-2g_{\mu\nu}\frac{\na_{\ga}\gO\na^{\ga}
\gO}{\gO^2}
\end{equation}
with $\gO^2=1+\ga(\stackrel{-}{\bx}\sqrt{\rho}/\sqrt{\rho})$.  Note that
the second relation in \eqref{9.4} is the Bohmian equation of motion and written
in terms of $g_{\mu\nu}$ it becomes $\na_{\mu}S\na^{\mu}S=m^2c^2$.
Many examples with a lot of expansion is to be found in \cite{c1,s2} and references
there.

\section{OTHER GEOMETRIC ASPECTS}
\renewcommand{\theequation}{4.\arabic{equation}}
\setcounter{equation}{0}

The quantum potential arises in many geometrical and cosmological situations and we mention a few of these here.
\begin{enumerate}
\item
We have written about the Wheeler-deWitt (WDW) equation and the QP in \cite{c17}
at some length and in \cite{c1} have discussed the QP in related geometric situations
following \cite{p3,p4,s5,s2,s19} in particular.  For background information on WDW
we refer to \cite{k9} for example.  One thinks of an ADM situation with $({\bf 4A})\,\,
ds^2=-(N^2-h^{ij}H_iN_j)dt^2+2N_idx^idt+h_{ij}dx^idx^j$ and the deWitt metric
$({\bf 4B})\,\,G_{ijk\ell}=(1/\sqrt{h})h_{ik}h_{j\ell}+h_{i\ell}h_{jk}-h_{ij}h_{k\ell})$.
Given a wave function $\psi=\sqrt{P}exp(iS/\hbar)$ where P can be thought of in terms of momentum fluctuations $(1/P)(\gd P/\gd h_{ij})$ one finds a quantum potential
\bq\label{4.1}
Q=-\frac{\hbar^2}{2}P^{-1/2}\frac{\gd}{\gd h_{ij}}\left(G_{ijk\ell}\frac{\gd P^{1/2}}{\gd
h_{k\ell}}\right)
\end{equation}
This is related to an intimate connection between Q and Fisher information 
based on techniques of Hall and Reginatto (cf. \cite{h2,h3,r1}.  The WDW equation
is
\bq\label{4.2}
\left[-\frac{\hbar^2}{2}\frac{\gd}{\gd h_{ij}}G_{ijk\ell}\frac{\gd}{\gd h_{k\ell}}+V\right]
\psi=0;
\end{equation}
and there is a lovely relation
\bq\label{4.3}
\int{\mf D}hPQ=-\int{\mf D}h\frac{\gd P^{1/2}}{\gd h_{ij}}
G_{ijk\ell}\frac{\gd P^{1/2}}{\gd h_{k\ell}}
\end{equation}
where the last term is Fisher information (cf. \cite{c17,f1,h2,h3,r1}).
\item
In \cite{s20} one uses again the attractive sandwich ordering of (4.2) (which is
mandatory in (4.2)) and considers WDW in the form
\bq\label{4.4}
\left[h^{-q}\frac{\gd}{\gd h_{ij}}h^qG_{ijk\ell}\frac{\gd}{\gd h_{k\ell}}+\sqrt{h}{}^{(3)}{\mc R}+\right.
\end{equation}
$$\left.+\frac{1}{2\sqrt{h}}\frac{\gd^2}{\gd \phi^2}-\frac{1}{2}\sqrt{h}h^{ij}\pp_i\phi\pp_j\phi
-\frac{1}{2}\sqrt{h}V(\phi)\right]\psi=0$$
with momentum constraint $({\bf 4C})\,\,i[2\na_j(\gd/\gd h_{ij})-h^{ij}\pp_j\phi(\gd/
\gd \phi)]\psi=0$ where $\phi$ is a matter field, $q$ is an ordering parameter, and 
$h=det(h_{ij})$.  Putting this in ``polar" form $\psi=\sqrt{\rho}exp(iS/\hbar)$ leads to
\bq\label{4.5}
G_{ijk\ell}\frac{\gd S}{\gd h_{ij}}\frac{\gd S}{\gd h_{k\ell}}+\frac{1}{2\sqrt{h}}
\left(\frac{\gd S}{\gd \phi}\right)^2-\sqrt{h}({}^{(3)}{\mc R}-{\mc Q}_G)+
\end{equation}
$$+\frac{\sqrt{h}}{2}h^{ij}\pp_i\phi\pp_j\phi+\frac{\sqrt{h}}{2}(V(\phi)-{\mc Q}_M)=0$$
where the gravity and matter quantum potentials are given via
\bq\label{4.6}
{\mc Q}_G=-\frac{1}{\sqrt{\rho h}}\left(G_{ijk\ell}\frac{\gd^2\sqrt{\rho}}{\gd h_{ij}\gd h_{k\ell}}
+h^{-q}\frac{\gd h^qG_{ijk\ell}}{\gd h_{ij}}\frac{\gd\sqrt{\rho}}{\gd h_{k\ell}}\right);\,\,
{\mc Q}_M=-\frac{1}{h\sqrt{\rho}}\frac{\gd^2\sqrt{\rho}}{\gd\phi^2}
\end{equation}
There is a continuity equation 
\bq\label{4.7}
\frac{\gd}{\gd h_{ij}}\left[2h^qG_{ijk\ell}\frac{\gd S}{\gd h_{k\ell}}\rho\right]+
\frac{\gd}{\gd\phi}\left[\frac{h^q}{\sqrt{h}}\frac{\gd S}{\gd\phi}\rho\right]=0
\end{equation}
and the momentum constraint leads to equations $({\bf 4D})\,\,2\na_j(\gd\sqrt{\rho}/\gd h_{ij})
-h^{ij}\pp_j\phi(\gd\sqrt{\rho}/\gd\phi)=0$ and $2\na_j(\gd S/\gd h_{ij})-h^{ij}\pp_j\phi(\gd S/\gd \rho)=0$ while the Bohmian ``guidance" equations are
\bq\label{4.8}
\frac{\gd S}{\gd h_{ij}}=\pi^{k\ell}=\sqrt{h}(K^{k\ell}-h^{k\ell}K);\,\,\frac{\gd S}{\gd \phi}=
\pi_{\phi}=\frac{\sqrt{h}}{N^{\perp}}\dot{\phi}-\sqrt{h}\frac{N^i}{N^{\perp}}\pp_i\phi
\end{equation}
where $K^{ij}$ is the extrinsic curvature.  Since in the WDW equation the wavefunction
is in the ground state with zero energy the stability condition of the metric and matter 
field is $({\bf 4E})\,\,h^{ij}\pp_i\phi\pp_j\phi+V(\phi)-2{}^3{\mc R}+{\mc Q}_M+2{\mc Q}_G=0$ which is a pure quantum solution (this follows from (4.5) by setting all
functional derivatives of S to be zero).   In \cite{s20} these equations are examined
perturbatively and 
we refer to \cite{s2,s16} for discussion of
the constraint algebra and related matters.
\item
In \cite{s16} one studies the constraint algebra and equations of motion based on a
Lagrangian $({\bf 4F})\,\,{\mf L}=\sqrt{-g}{\mc R}=\sqrt{h}N({}^{(3)}{\mc R}+
Tr(K^2)-(Tr K)^2)$ where ${}^{(3)}$is the 3-D Ricci scalar, $K_{ij}$ the extrinsic
curvature, and $h$ the induced spatial metric.  The canonical momentum of the 
3-metric is given via $({\bf 4G})\,\,P^{IJ}=\pp{\mf L}/\pp \dot{h}_{ij})=\sqrt{h}(K^{ij}-h^{ij}
Tr K)$ and the classical Hamiltonian is $({\bf 4H})\,\,H=\int d^3x{\mf H}$ with ${mf H}=\sqrt{h}(NC+N^iC_i)$.  Here one has
\bq\label{4.9}
C=-{}^{(3)}{\mc R}+\frac{1}{h}\left(Tr(p^2)-\frac{1}{2}(Tr p)^2)\right)=-2G_{\mu\nu}n^{mu}
n^{\nu};
\end{equation}
$$C_i=-2{}^{(3)}\na^j\left(\frac{p_{ij}}{\sqrt{h}}\right)=-2G_{\mu i}n^{\mu}$$
where $n^{\mu}$ is normal given via $n^{\mu}=(1/N,-{\bf N}/N)$.  To get the quantum
version one takes $H\to H+Q$ (${\mf H}\to {\mf H}+{\mc Q}$ (where $Q=\int d^3x
{\mc Q}$) and
\bq\label{4.10}
{\mc Q}=\hbar^2NhG_{ijk\ell}\frac{1}{|\psi|}\frac{\gd^2|\psi|}{\gd h_{ij}\gd h_{k\ell}}
\end{equation}
The classical constraints are then modified via $C\to C+({\mc Q}/\sqrt{h}N)$ and
$C_i\to C_i$.  We disregard the constraint algebra here and go some formulas for
quantum Einstein equations.  First there is an HJ equation
\bq\label{4.11}
G_{ijk\ell}\frac{\gd S}{\gd h_{ij}}\frac{\gd S}{\gd h_{k\ell}}-\sqrt{h}({}^{(3)}{\mc R}
-{\mc Q})=0
\end{equation}
where S is the phase of the wave function and this leads to Bohm-Einstein equations
\bq\label{4.12}
{\mc G}^{ij}=-\gk{\mc T}^{ij}-\frac{1}{N}\frac{\gd({\mc Q}_G+{\mc Q}_m}{\gd g_{ij}};\,\,
{\mc G}^{0\mu}=-\gk{\mc T}^{0\mu}+\frac{{\mc Q}_G+{\mc Q}_m}{2\sqrt{-g}}g^{0\mu};
\end{equation}
$${\mc Q}_m=\hbar^2\frac{N\sqrt{H}}{2}\frac{\gd^2|psi|}{\gd\phi^2};\,\,{\mc Q}_G=
\hbar^2NhG_{ijk\ell}\frac{1}{|\psi|}\frac{\gd^2|\psi|}{\gd h_{ij}\gd h_{k\ell}}$$
These are the quantum version of the Einstein equations and since regularization
here only affects the quantum potential (cf. \cite{s16}) for any regularization the
quantum Einstein equations are the same and one can write
$({\bf 4I})\,\,{\mc G}^{\mu\nu}=-\gk{\mc T}^{\mu\nu}+{\mf S}^{mu\nu}$ with
\bq\label{4.13}
{\mf S}^{0\mu}=-\frac{{\mc Q}_G+{\mc Q}_m}{2\sqrt{-g}}g^{0\mu}=\frac{{\mc Q}}
{2\sqrt{-g}}g^{0\mu};\,\,{\mf S}^{ij}=-\frac{1}{N}\frac{\gd Q}{\gd g_{ij}}
\end{equation}
\item
There are also developments of Bohmian theory and quantum geometrodynamics
in \cite{b22,b23,b20,c22,k10,p3,p4,s5,s21} (cf. \cite{c1} for some survey and more references).  In 
\cite{p3} for example one writes the WDW equation in the form
\bq\label{4.14}
\left\{-\hbar^2\left[\gk G_{ijk\ell}\frac{\gd}{\gd h_{ij}}\frac{\gd}{\gd h_{k\ell}}+\frac{1}{2}
h^{-1/2}\frac{\gd^2}{\gd\phi^2}\right]+\right\}\psi(h_{ij},\phi)=0;
\end{equation}
$$V=h^{1/2}\left[-\gk^{-1}({\mc R}^{(3)}-2\gL)+\frac{1}{2}h^{ij}\pp_i\phi\pp_j\phi+U(\phi)
\right]$$
(questions of factor ordering and regularization are ignored here) with a constraint  
$({\bf 4J})\,\,-2h_{ij}\na_j(\gd \psi/\gd h_{ij})+(\gd \psi)/\gd\phi)\pp_i
\phi=0$.
Writing now
$\psi=Rexp(iS/\hbar)$ ({\bf 4J}) leads to
\bq\label{4.15}
-2h_{ij}\na_j(\gd S/\gd h_{ij})+(\gd S/\gd\phi)\pp_i\phi=0;\,\,-2h_{ij}\na_j(\gd R/\gd h_{ij})
+(\gd R/\gd \phi)\pp_i\phi=0
\end{equation}
and (4.14) yields
\bq\label{4.16}
\gk G_{ijk\ell}\frac{\gd S}{\gd h_{ij}}\frac{\gd S}{\gd h_{k\ell}}+\frac{1}{2}h^{-1/2}
\left(\frac{\gd S}{\gd\phi}\right)^2+V+Q=0;
\end{equation}
$$Q=-\frac{\hbar^2}{R}\left(\gk G_{ijk\ell}\frac{\gd^2R}{\gd h_{ij}h_{j\ell}}+
\frac{h^{-1/2}}{2}\frac{\gd^2R}{\gd \phi^2}\right);$$
$$\gk G_{ijk\ell}\frac{\gd}{\gd h_{ij}}\left(R^2\frac{\gd S}{\gd h_{k\ell}}\right)+\frac
{1}{2}\frac{\gd}{\gd \phi}\left(R^2\frac{\gd S}{\gd\phi}\right)=0$$
\item
In \cite{s4} one picks up again the approach of (2) to find a pure quantum state
leading to a static Einstein universe whose classical counterpart is flat spacetime.
For WDW one uses a form of (4.4) (with $16\pi G=1$ and ${\mc R}$ the 3-curvature
scalar), namely
\bq\label{4.17}
\hbar^2h^{-q}\frac{\gd}{\gd h_{ij}}\left(h^qG_{ijk\ell}\frac{\gd \psi}{\gd h_{k\ell}}\right)
+\sqrt{h}{\mc R}\psi+\frac{1}{\sqrt{h}}{\mc T}^{00}\left(\frac{-i\hbar\gd}{\gd\phi_a},
\phi_a\right)\psi=0;
\end{equation}
with 3-diffeomorphism constraint in the form $({\bf 4K})\,\,2\na_j(\gd/\gd h_{ij})\psi-{\mc T}^{i0}(\gd/\gd\phi_a,\phi_a)\pi=0$.  ${\mc T}^{\mu\nu}$ is the energy momentum tensor of matter fields $\phi_a$ in which the matter is quantized by
replacing its conjugate momenta by $-i\hbar\gd/\gd\phi_a$.  For the causal
interpretation one sets again $\psi=Rexp(iS/\hbar)$ to obtain
\bq\label{4.18}
G_{ijk\ell}\frac{\gd S}{\gd h_{ij}}\frac{\gd S}{\gd h_{k\ell}}-\sqrt{h}({\mc R}-Q_G)
+\frac{1}{\sqrt{h}}({\mc T}^{00}(\gd S/\gd \phi_a,\phi_a)+Q_M)=0;
\end{equation}
\bq\label{4.19}
\frac{\gd}{\gd h_{ij}}\left(2h^qG_{ijk\ell}\frac{\gd S}{\gd h_{k\ell}}R^2\right)+
\sum\frac{\gd}{\gd \phi_a}\left(h^{q-(1/2)}\frac{\gd S}{\gd \phi_a}R^2\right)=0
\end{equation}
\bq\label{4.20}
Q_G=-\frac{\hbar^2}{\sqrt{h}R}\left(h^{-q}\frac{\gd}{\gd h_{ij}}h^qG_{ijk\ell}\frac
{\gd R}{\gd h_{k\ell}}\right);\,\,Q_M-\frac{\hbar^2}{hR}\sum\frac{\gd^2R}{\gd\phi_a^2}
\end{equation}
\bq\label{4.21}
2\na_j\frac{\gd R}{\gd h_{ij}}-{\mc T}^{i0}(\gd R/\gd\phi_a,\phi_a)=0;\,\,2\na_j
\frac{\gd S}{\gd h_{ij}}-{\mc T}^{i0}(\gd S/\gd\phi_a,\phi_a)=0
\end{equation}
One notes that all terms containing the second functional derivative are ill defined
and can be regulated via $(\gd/\gd h_{ij}(x))(\gd/\gd h_{ij}(x)\to \int d^3x\sqrt{h}
U(x-x')(\gd/\gd h_{ij}(x))(\gd/\gd h_{ij}(x'))$ where U is the regulator.  Finally the guidance equations are $({\bf 4L})\,\,\pi^{k\ell}=\sqrt{h}(K^{k\ell}-Kh^{k\ell})=
\gd S/\gd h_{k\ell}$ and $\pi_{\phi_a}=\gd S/\gd\phi_a$ where $K_{ij}=(1/2N)
(\dot{h}_{ij}-\na_iN_j-\na_jN_i)$ is the extrinsic curvature.
Using the quantum  Hamilton-Jacobi-Einstein (HJE) equation one can define a
limit, called the pure quantum limit, where the total quantum potential is of the same
order as the total classical potential and they can cancel each other.  
In this case one has $({\bf 4M})\,\,\gd S/
\gd h_{ij}=\gd S/\gd \phi_a=0$ and the continuity is satisfied identically.  The
resulting trajectory is not similar to any classical solution and the quantum HJE
equation for a pure quantum state is an equation for spatial dependence of the metric and matter fields in terms of the norm of the wave function.
Explicit calculations are given for some special situations.
\end{enumerate}
\indent
{\bf REMARK 4.1.}
Note that for $\eta_{ab}\sim (1,-1,-1,-1)$ and $\hbar=c=1$ one has $\pp_0^2-\na^2\sim\bx$ and 
$({\bf 4N})\,\,(\na S)^2=m^2[1+(\bx R/m^2R)]$ as in (3.4) (Nikoli'c).  This agrees with
\bq\label{4.22}
(\na S)^2={\mf M}^2c^2(1+Q);\,\,Q=\frac{\hbar^2}{m^2c^2}\frac{\bx R}{R}
\end{equation}
from Section 3.1 (F. and A. Shojai).  For the BFM theory with $\eta_{ab}\sim
(-1,1,1,1)$ one has 
$({\bf 4O})\,\,(1/2m)(\na S)^2+(mc^2/2)-(\hbar^2/2m)(\bx R/R)=0$
from (3.2).  But $\bx R\to -\bx R$
and $(\na S)^2\sim\eta^{ab}\na_bS\na_aS\to -(\na S)^2$ for $\eta_{ab}\to-\eta_{ab}$.
Hence in the $\eta_{ab}=(1,-1,-1,-1)$ notation one obtains
$(\na S)^2=m^2c^2[1+(\hbar^2/m^2c^2)(\bx R/R)]$ as in (4.22).$\hfill\bs$
\\[3mm]\indent
{\bf REMARK 4.2.}
There is a lot of motivation here for using the quantum potential as a generator of
quantum gravity (cf. \cite{b24}) and also for considering the conformal factor
${\mf M}^2/m^2$ as a generator of Ricci flow (cf. \cite{c23,g13,p11}).
$\hfill\bs$
\\[3mm]\indent
{\bf REMARK 4.3.}
One finds fascinating connections between Bohmian theory and
phase space mechanics in \cite{b3,b5,g1,g14,g12,n5,n10,n11,s22}.
In \cite{b3} one argues that if the quantum
potential (QP) reflects the quantum aspects of a sysem it should be possible to to identify such
aspects within the QP and in particular one shows how the balance between localisation and
dispersion energies suggests a link between the QP and the Heisenberg uncertainty principle.
Recall first that from the SE $({\bf 5A})\,\,i\hbar\pp_t\psi=[-(\hbar^2/2m)\na^2+V]\psi$ there
follows
\bq\label{03.1}
\pp_tS+\frac{(\na
S)^2}{2m}-\frac{\hbar^2}{2m}\frac{\na^2R}{r}+V=0;\,\,\pp_t\rho+\na\cdot\left(\rho\frac{p}{m}\right)=0
\end{equation}
where $\psi=Rexp(iS/\hbar),\,\,\rho=|\psi|^2,$ and $p=\na S$.  The QP is manifestly of the form
$Q=-(\hbar^2/2m)(\na^2R/R)$ and one writes $F=-\na(Q+V)$ and $v=j/\rho=p/m$.  Now consider a more
general derivation of the QP by writing the SE in the form
$({\bf 5B})\,\,i\hbar\pp_t\psi=(T(\hat{p})+V(\hat{x}))\psi$.  Setting again $\psi=Rexp(iS/\hbar)$ one
obtains
\bq\label{03.2}
\pp_tS+\Re\left(\frac{T\psi}{\psi}\right)+V(x)=0;\,\,\pp_t\rho-\frac{2\rho}{\hbar}\Im\left(\frac
{T\psi}{\psi}\right)=0
\end{equation}
Correspondingly in the momentum space with $\hat{x}=i\hbar\na_p$ and $\hat{p}=p$ the real and
imaginary parts of the SE are
\bq\label{03.3}
\pp_tS+T(p)+\Re\left(\frac{V\psi}{\psi}\right)=0;\,\,\pp_t\rho-\frac{2\rho}{\hbar}\Im\left(\frac
{V\psi}{\psi}\right)=0
\end{equation}
Then expanding exponentials one writes
\bq\label{03.4}
\Re\left(\frac{\psi^*T\psi}{\rho}\right)=\Re\left(\frac{R[1-(iS/\hbar)-\cdots)T(\hat{p})R(1+(iS/\hbar)-
\cdots]}{\rho}\right)
\end{equation}
(note the formal equivalence $T\psi/\psi=\psi^*T\psi/\rho$).  If now $T(\hat{p})$ is a general but
analytic function of $\hat{p}$ one can expand in a power series in $\hat{p}=-i\hbar\na$ and
the kinetic term may be separated into the sum of two parts
\bq\label{03.5}
\Re\left(\frac{T\psi}{\psi}\right)=T_h(x)+T_0(x);\,\,T_0(x)=T(\na S)
\end{equation}
where $T_h(x)$ is an expansion in even positive powers of $\hbar$ and $T_0(x)$ is independent of
$\hbar$ and identifies $p=\na S$.  The same line of argument allows the potential term of the HJ
equation in \eqref{03.3} to be separated as
\bq\label{03.6}
\Re\left(\frac{V\psi}{\psi}\right)=V_h(p)+V_0(p);\,\,V_0(p)=V(-\na_pS)
\end{equation}
where $V_h$ is an expansion in even positive powers of $\hbar$ and $V_0(p)$ is independent of
$\hbar$ and identifies $x=-\na_pS$.
We pursue this further in \cite{c25}.$\hfill\bs$

\section{THE QUANTUM POTENTIAL AND GEOMETRY}
\renewcommand{\theequation}{5.\arabic{equation}}
\setcounter{equation}{0}

We begin with \cite{s2} and recall some features of Weyl geometry
(some of which are indicated already in previous sections).  We remember first
that vectors change in length and direction under translation via $({\bf 5A})
\,\,\gd\ell=\phi_{\mu}\gd x^{\mu}\ell$ so $\ell=\ell_0exp(\int \phi_{\mu}dx^{\mu})$ where $\phi_{\mu}$ is the Weyl vector.  Equivalently $({\bf 5B})\,\,
g_{\mu\nu}\to exp(2\int\phi_{\mu}\gd x^{\mu})g_{\mu\nu}$ which is a conformal transformation.  Recall also that the metric is a Weyl covariant
object of weight 2 and the Weyl connection is given via
\bq\label{5.1}
\gG^{\mu}_{\nu\gl}=\left\{\begin{array}{c}
\mu\\
\nu\gl\end{array}\right\} +g_{\nu\gl}\phi^{\mu}-\gd^{\mu}_{\nu}\phi_{\gl}
-\gd_{\gl}^{\mu}\phi_{\nu}
\end{equation}
A gauge transformation $({\bf 5C})\,\,\phi_{\mu}\to\phi'_{\mu}=\phi_{\mu}
+\pp_{\mu}\gL$ transforms 
$g_{\mu\nu}\to g'_{\mu\nu}=exp(2\gL)g_{\mu\nu}$ with $\gd\ell\to\gd\ell'=\gd\ell+(\pp_{\mu}\gL)dx^{\mu}\ell$.
In remarks just before Section 3.1 we have seen how quantum effects are
geometrized via the Dirac field $\gb$ and gauge transformations.  Let us
now make more explicit some direct relations between the quantum potential and geometric ideas via the Weyl vector.  We recall from Section 3
($\# 2$) that for the SE $Q=-(m/2){\bf u}^2-(\hbar/2)\pp{\bf u}$ where
${\bf u}$ is an osmotic velocity (see also \cite{c1,g17}).  Similarly for the KG
equation one has $Q=(\hbar^2/m^2c^2)(\bx_g|\psi|/|\psi|$ for example
as in Remark 4.1 and we will also consider an appropriate osmotic velocity for
this situation.
\\[3mm]\indent
Consider now the situation $Q=0$ for the SE which can be expressed in several forms.
\begin{enumerate}
\item
Defining the osmotic velocity as ${\bf u}=D\na log(\rho)$ with $D=\hbar/2m$ from 
\cite{c1,g17} (cf. also \cite{n4,n8}) one has then $({\bf 5A})\,\,(m/2){\bf u}^2+
(\hbar/2)\na {\bf u}=0$.
\item
Another form is directly (for $g=1$) $({\bf 5B})\,\,\gD\sqrt{\rho}=0$.
\item
There is a general form for $({\bf 5C})\,\,\phi_i=-\pp_ilog(\hat{\rho})$ with
$\hat{\rho}=\rho/\sqrt{g}$, namely
\bq\label{5.2}
\dot{{\mc R}}+2\left[\phi_i\phi^i-\frac{2}{\sqrt{g}}\pp_i(\sqrt{g}\phi^i)\right]=0
\end{equation}
When $\dot{{\mc R}}=0$ with $\sqrt{g}=1$ and $\hat{\rho}=\rho$ this becomes
$({\bf 5D})\,\,\phi_i\phi^i-2\pp_i\phi^i=0$.  Note $\phi^i\sim g^{ik}\phi_k=-g^{ik}\pp_k
log(\hat{\rho})=-\pp^klog(\hat{\rho})$.
\item
From (2.8) another form of (5.2) above is
\bq\label{5.3}
\dot{{\mc R}}+\frac{8}{\sqrt{\rho}}\pp_i(\sqrt{g}g^{ik}\pp_k\sqrt{\hat{\rho}})=0
\end{equation}
\item
In view of \cite{c1,c5,c23,h2,h3,r1} one can say that the fundamental quantum
fluctuation or perturbation in momentum has the form $({\bf 5E})\,\,\gd p\sim
c(\na\rho/\rho)$ and this means $({\bf 5F})\,\,\gd p\sim \hat{c}{\bf u}$ or equivalently
$\gd p\sim\tl{c}\vec{\phi}$.  We can assume that in a Weyl space situation an osmotic
velocity ${\bf u}=Dlog(\hat{\rho})$ is meaningful.  The ``obligatory" nature of $\gd p\sim
c(\na\rho/\rho)$ is made even more striking in the developments in \cite{h31,m2}.
One shows there in particular that a classical momentum can be written as 
\bq\label{5.33}
\hat{p}_{cl}=\hat{p}+\left(\frac{i\hbar}{2}\right)\left(\frac{\na\rho}{\rho}\right)\Rightarrow
\hat{p}_{cl}=-i\hbar\left(\na-\frac{1}{2}\frac{\na(\psi^*\psi)}{\psi^*\psi}\right)
\end{equation}
\end{enumerate}
\indent
It should now be possible to extract some analytic and geometric features of the situation
$Q=0$.
\begin{example}
We think of $\psi=\sqrt{\rho}exp(iS/\hbar)$ with $\sqrt{\rho}=R$.  Take $\# 2$ first and look for solutions of $\gD R=0$ in a finite region $\gO$ with $R\in H_0^1(\gO)$ (Sobolev space) for example (see \cite{c26,e1} for techniques and results in PDE).  For a 
QM situation $R=0$ on $\pp\gO$ and $H^1_0(\gO)$ is the natural setting with 
$R \in L^2(\gO)$.  However by Green's theorem $\int_{\gO}R\gD RdV=-\int_{\gO}
|\na R|^2dS=0$ which implies $\na R=R=0$.  this is consistent with the Example 1.2
where $\psi$ involves plane waves and $L^2$ solutions are meaningless.$\hfill\bs$
\end{example}
\begin{example}
Consider next a situation $(m/2)|{\bf u}|^2+(\hbar/2)\na{\bf u}=0$ or in 1-D $\pp u+cu^2=0$ with $c>0$.  Then $u'/u^2=-c\Rightarrow u=(\hat{c}+ cx)^{-1}$ and setting $u=D\rho'/
\rho$ yields $R^2=\rho=k(\hat{c}+cx)^{d}$.  This is not reasonable for $R=0$ outside
of a finite $\gO$.$\hfill\bs$
\end{example}
\begin{example}
Consider $\phi_i\phi^i-2\pp\phi^i=0$ or equivalently (in 1-D for convenience) $\phi'/\phi^2
=1/2$ leading to $-(1/\phi)=(1/2)x+c$ and problems similar to those in Example 5.2.
$\hfill\bs$
\end{example}
\indent
We can however think of $\rho,\,\vec{\phi},$ or ${\bf u}$ as functions of Q so for each admisible
Q there will be in principle some well determined R, modulo spectral conditions as
in Remark 1.1.
\\[3mm]\indent
{\bf REMARK 5.1.}
In Remark 1.1 we saw that determining R from Q involved solving $\gD R
+\gb QR=0\,\,(\gb>0)$ in say $H_0^1(\gO)$.  If $Q\leq 0$ this yields a unique
solution while if $0$ is not in the spectrum of $\gD +\gb Q$ then 
$({\bf 5G})\,\,\gD R+\gb
QR=0$ has a unique solution for say $Q\in L^{\infty}(\gO)$.  We also
saw that modulo solvability of (1.4) one would obtain a ``generalized" 
quantum theory based on Q.  We can improve the statement of this in
Remark 1.1 by saying that, given solutions V and S of (1.4) (via a solution R
of ({\bf 5G})), in converting this to a SE one eliminates Q from the picture
entirely.  For R unique $S(x,t)$ is determined up to a function $f(t)$ 
and a function $g(x)$ arising from
\bq\label{5.4}
S=-\int^t(Q+V)dt-\frac{1}{2R^4}\left[f(t)-\int^x\pp_tR^2dx\right]^2+g(x)
\end{equation}
However $V_x$ is known via (1.4) in terms of $S_{xt}$ which depends only on Q
(via R), $f$, and $f'$, hence only in terms of one function $f(t)$.
If then $V=V(x)$ it may actually be almost determined and
Q, instead of determining only one trajectory based on R and S,
actually could lead to the SE itself (modulo $f$) for $\psi=Rexp(iS/\hbar)$;
if it were to be the case that $V=V(x)$ does not use $f(t)$ this means
that Q alone would determine a ``generalized" quantum theory via the SE!  
This could eliminate some of the ambiguity connected with the idea
of using Q as a quantization.  It would be worthwhile checking the equations to
find such situations (see below).  We note also that if Q contains $t$ it is transmitted to R as a
parameter in solving the elliptic equation; if Q is independent of $t$ then of course
so is R and this could conceivably simplify matters in determining $V=V(x)$.
$\hfill\bs$
\\[3mm]\indent
We check this last idea in more detail now.  Assume Q is a function of $x$ alone,
$Q=Q(x)$, and let it determine a unique $R\in H_0^1(\gO)$ (normalized so that
$\int_{\gO}R^2dx=1$).  Then look at (1.4), namely
\bq\label{5.5}
S_t+\frac{1}{2m}S_x^2+Q+V=0;\,\,\pp_tR^2+\frac{1}{m}(R^2S_x)_x=0
\end{equation}
The second equation becomes $(R^2S_x)_x=0$ which implies $({\bf 5H})\,\,R^2S_x=
f(t)$ for some ``arbitrary" $f(t)$.  Then $({\bf 5I})\,\,S_t+(1/2m)(f^2/R^4)+Q+V=0$ and 
we can eliminate S from ({\bf 5H}) and ({\bf 5I}) via
\bq\label{5.6}
R^2S_{xt}=f_t;\,\,S_{xt}-\frac{2f^2R_x}{mR^5}+Q_x+V_x=0\Rightarrow
\frac{2f^2}{m}\frac{R_x}{R^3}-f_t=R^2(Q_x+V_x)
\end{equation}
This determines $V_x$ in terms of $Q(x)$ and $f(t)$ so we ask whether $V=V(x)$
can occur (no $t$ dependence).  In such a case the $t$ derivatives of the last term
in (5.7) are zero yielding $({\bf 5J})\,\,f_{tt}=(2R_x/mR^3)\pp_tf^2$.
This means
\bq\label{5.8}
\frac{f_{tt}}{\pp_tf^2}=F(t)=\frac{2R_x}{mR}={\mf F}(x)
\end{equation}
Consequently $F(t)={\mf F}(x)=c$ and $({\bf 5K})\,\,f_{tt}=c\pp_tf^2$ while $(R_x/R^3)=(cm/2)$ leading to
\bq\label{5.9}
f_t=cf^2+\hat{c};\,\,R^2=\frac{1}{\tl{c}-cmx}
\end{equation} 
Thus $x>(\tl{c}/cm)$ but $R\notin H_0^1(\gO)$ for any $\gO$.  This seem to preclude $V=V(x)$ (or perhaps $Q=Q(x)$).
Hence from (5.6) one has at least (cf. also Section 9)
\begin{proposition}
Given $Q=Q(x)$ determining a unique $R(x)\in H_0^1(\gO)$ it follows that $V_x$
is determined up to an ``arbitrary" function $f(t)$ via
\bq\label{5.9}
V_x=\frac{1}{R^2}\left[\frac{2f^2R_x}{mR^3}-f_t\right]-Q_x
\end{equation}
This situation precludes $V$ being a function of $x$ alone.$\hfill\bs$
\end{proposition}
\indent
{\bf REMARK 5.2.}
Note if $\int_{\gO}R^2(x,t)dx=r^2(t)$ then to get a proper normalization one would take ${\mc R}(x,t)=(1/r(t))R(x,t)$ and note that Q computed on ${\mc R}$ is equal via
$({\bf 5L})\,\,{\mc Q}=-(\hbar^2/2m)({\mc R}_{xx}/{\mc R})=-(\hbar^2/2m)
(R_{xx}/R)$.  Note also that $r$ is determined by Q via R.  We still think of $\psi\sim
Rexp(iS/\hbar)$ so (5.5) applies and ({\bf 5H}) becomes $({\bf 5M})\,\,
RS^2_x=-m\int^x\pp_tR^2dx+f(t)=A(f,Q)$ (since Q determines R).  Then we are
still essentially in the context of Remark 5.1 and $({\bf 5N})\,\,2RR_tS_x+
R^2S_{xt}=\pp_tA$ while from (5.5) $({\bf 5O})\,\,S_{xt}+(1/m)S_xS_{xx}+Q_x
+V_x=0$.  Now we eliminate $S_{xx}$ and $S_{xt}$ to get $V_x$ in terms of Q and
$f$.  First from ({\bf 5N}) one has
\bq\label{5.10}
S_{xt}=\frac{1}{R^2}\left\{\pp_tA-2\frac{R_t}{R}A\right\}
\end{equation}
while $({\bf 5P})\,\,R^2S_{xx}+2RR_x=\pp_xA\Rightarrow S_{xx}=(1/R^2)
(\pp_xA-2RR_x)$.  Hence one arrives at
\bq\label{5.11}
\frac{1}{R^2}\left[\pp_tA-\frac{2AR_t}{R}\right]=-Q_x-V_x-\frac{1}{m}\frac{A}{R^4}
(\pp_xA-2RR_x)
\end{equation}
and we can state
\begin{proposition}
Defining $A(f,Q)=f(t)=m\int^x\pp_tR^2dx$ with $f$ ``arbitrary" one can determine
$V_x$ via (5.12) as $V_x(Q,f)$.  Hence $V=\int^xV_xdx+h(t)$ for $h$ ``arbitrary" 
provides a potential $V(Q,f,h)$ and the associated SE is determined completely by
V.  If choices $f=h=0$ are ``natural" one can say that Q determines a natural SE
and a corresponding ``generalized" quantum theory.$\hfill\bs$
\end{proposition}
\indent
{\bf REMARK 5.3.}
Consider the stationary case (cf. \cite{c1}) $({\bf 5Q})\,\,(1/2m)S_x^2+Q+V-E=0$
with $(R^2S_x)_x=0$ where $R=R(x)$ is say uniquely determined via $Q=
Q(x)$ (note however that both R and Q must contain E as a parameter).  Then
\bq\label{5.12}
S_x=\frac{c}{R^2};\,\,\frac{1}{2m}\left(\frac{c^2}{R^4}\right)+Q+V-E=0
\end{equation}
This means $({\bf 5R})\,\,1=\pp_EQ-(c^2/mR^5)\pp_ER$
(since V does not depend on E) and hence
\bq\label{5.13}
\frac{2c^2R_E}{mR^5}+1=\frac{\hbar^2R''R_E}{2mR^2}-\frac{\hbar^2R_E''}{2mR}
\end{equation}
Viewed in terms of $\rho=R^2$ this mean that $\rho=\rho(E,x)$ and the corresponding Weyl geometry based on $\vec{\phi}=-\na log(\rho)$  
will depend on E (as will Q of course).  We refer here also to the quantum mass
idea of Floyd, namely $m_Q=m(1-\pp_EQ)$ for stationary situations (this is 
sketched in \cite{c27} for example and we refer to \cite{f10} for more details).
Some further ideas about this are sketched in Remark 7.3.  In particular one knows
that $Q\sim -(\hbar^2/2m){\mc R}$ (from Section 2) where ${\mc R}$ is the 
Ricci-Weyl curvature with $\hbar$ essentially put in by hand to conform to the 
wave function idea and operator QM.  We see that the geometry of the space 
in which a trajectory transpires is thereby determined by E (not surprisingly)
which seems to say that the probability distribution $\rho$ is the basic unknown
here (and in the time dependent situation).  Once one has a probability distribution
one can posit a wave function and insert $\hbar$.  In fact (given V) the two equations
$(1/2m)(c^2/R^2)+Q+V-E=0$ and $Q=-(\hbar^2/2m)(R''/R)$ determine $R=R(E,x)$
directly ($\hbar$ being gratuitously inserted).
$\hfill\bs$

\section{OLAVO THEORY}
\renewcommand{\theequation}{6.\arabic{equation}}
\setcounter{equation}{0}

We go here to \cite{o1} and sketch some matters dealing with uncertainty and the
SE (cf. also \cite{b2,c27}).  
In the first paper of \cite{o1} an axiomatic formulation for quantum mechanics (QM)
is given.  Consider ensembles described by probability density
functions in phase space described as $F(x,p,t)$; assume
\begin{enumerate}
\item
Newtonian particle mechanics is valid for particles in the ensemble.
\item
For an isolated system  $dF(x,p,t)/dt=0$.
\item
The Wigner-Moyal infinitesimal transformation is defined via
\bq\label{6.1}
Z_Q(x,\gd x/2,t)=\int F(x,p,t)exp\left(\frac{ip\gd x}{\ell}\right)dp
\end{equation}
where $\ell$ is a parameter which will necessarily be equal to $\hbar$.
\end{enumerate}
From these axioms one can derive nonrelativistic QM as follows.
First using (1) and (2) one has
\bq\label{6.2}
\frac{dF}{dt}=\pp_tF+\dot{x}F_x+\dot{p}F_p=0;\,\,\dot{x}=\frac{p}{m};\,\,\dot{p}=f=-V_x
\end{equation}
Multiplying by the exponential in (6.1) and integrating one arrives at
\bq\label{6.3}
-\pp_tZ_Q+\frac{i\hbar}{m}\frac{\pp^2Z_Q}{\pp x\pp(\gd x)}-\frac{i}{\hbar}\gd V(x)Z_Q=0
\end{equation}
where the infinitesimal nature of $\gd x$ is used to write
\bq\label{6.4}
\pp_xV\gd x=\gd V(x)=V\left(x+\frac{\gd x}{2}\right)-V\left(x-\frac{\gd x}{2}\right);
\end{equation}
and one knows that $(\bgs)\,\,
F(x,p,t)exp\left(\frac{ip\gd x}{\ell}\right)^{p=\infty}_{p=-\infty}=0$ by the nature
of probability distributions.
Writing $y=x+(\gd x/2)$ and $y'=x-(\gd x/2)$ one can rewrite (6.3) as
\bq\label{6.5}
\left\{\frac{\hbar^2}{2m}[\pp_y^2-\pp_{y'}^2]-[V(y)-V(y')]\right\}Z_Q(y,y',t)=-i\hbar\pp_t
Z_Q(yy',t)
\end{equation}
This is called a Schr\"odinger equation (SE) for the characteristic function $Z_Q$
and is valid for all values of $y,\,y'$ as long as they are infinitesimally close.
\\[3mm]\indent
Now suppose one can write (this is a fundamental assumption)
\bq\label{6.6}
Z_Q(y,y',t)=\psi^*(y',t)\psi(y,t);\,\,\psi(y,t)=R(t,y)e^{iS(y,t)/\hbar}
\end{equation}
Then expanding $Z_Q$ one obtains
\bq\label{6.7}
Z_Q(y,y',t)=\left\{R(x,t)^2-\left(\frac{\gd x}{2}\right)^2\left[(\pp_xR)^2-R\frac{\pp^2R}{\pp x^2}
\right]\right\}exp\left(\frac{i}{\hbar}(\gd x)S_x\right)
\end{equation}
Putting this in (6.3) leads to
\bq\label{6.8}
P_t+\pp_x\left(\frac{PS_x}{m}\right)=0;\,\,\frac{i\gd x}{\hbar}\pp_x\left[
\frac{S_x^2}{2m}+V+S_t-\frac{\hbar^2}{2mR}\frac{\pp^2R}{\pp x^2}\right]=0
\end{equation}
where $P(x,t)=R^2=lim_{\gd x\to 0}Z_Q(x+(\gd x/2),x-(\gd x/2),t)$ is the probability distribution in configuration space. Equation 
(6.8b) can be rewritten as 
\bq\label{6.9}
\frac{1}{2m}S_x^2+V+S_t+Q=f(t);\,\,Q=-\frac{\hbar^2}{2mR}\pp^2_xR
\end{equation}
One can eliminate $f(t)$ by redefining $S(x,t)$ as $S'(x,t)=S(x,t)+\int_0^tf(t')dt'$;
since this is only a new definition of the energy reference level one can simply take 
$f(t)=0$ and obtain as well $({\bf 6A})\,\,(\hbar^2/2m)\psi_{xx}-V(x)\psi=-i\hbar\psi_t$.
Thus if one can write $Z_Q$ as the product (6.6) then $\psi$ satisfies the SE ({\bf 6A})
and (6.8) holds.  
\\[3mm]\indent
Now define operators on $Z_Q$ via primes (distinguished from operators acting on the probability amplitude) so that $\hat{p}'=-i\hbar(\pp/\pp(\gd x))$
and $\hat{x}'=x$; this is based on the fact that
\bq\label{6.10}
\bar{p}=\lim_{\gd x\to 0}-i\hbar\frac{\pp}{\pp(\gd x)}\int F(x,p,t)exp\left(i\frac{p\gd x}{\hbar}\right)dxdp
\end{equation}
and $\bar{x}=lim_{\gd x\to 0}\int xF(x,p,t)exp[ip\gd x/\hbar]dxdp$.  Thus the result of
position and momentum operators acting on $Z_Q$ represents a mean value
calculation for the ensemble components.   Observe now via (6.7) that
\bq\label{6.11}
\bar{p}=lim_{\gd x\to 0}\left[-i\hbar\frac{\pp}{\pp(\gd x)}\int F(x,p,t)exp\left(i\frac{p\gd x}{\hbar}
\right)dxdp\right]=\int R^2(x)S_xdx
\end{equation}
One can rewrite (6.11) in the form
\bq\label{6.12}
\hat{p}=lim_{\gd x\to 0}\left\{-i\hbar\int\frac{\pp}{\pp(\gd x)}\left[\psi^*\left(x-\frac{\gd x}{2},t
\right)\psi\left(x+\frac{\gd x}{2},t\right)\right]dx\right\}
\end{equation}
Again via (6.7) this leads to $({\bf 6B})\,\,\hat{p}=\int \psi^*x,t)(-i\hbar\pp_x)\psi(x,t)dx$
with similar calculations for the position operator.  Consequently $({\bf 6C})\,\,
\hat{p}\psi=-i\hbar\pp_x\psi$ and $\hat{x}\psi=x\psi$.  Moreover some calculation 
shows that $[\hat{x},\hat{p}]=i\hbar$ and consequently $\gD x\gD p\geq \hbar/2$
(cf. \cite{o1} for details).
\\[3mm]\indent
Next we go to the third paper in \cite{o1} to sketch another derivation of the SE
from more physically grounded axioms and then the derivations are connected.  Thus begin
with the Liouville equation for $F(x,p,t)$
\bq\label{6.13}
\pp_tF+\frac{p}{m}\pp_xF-\pp_xV\pp_pF=0
\end{equation}
Integrating in $p$ and using the definitions $({\bf 6D})\,\,\int Fdp=\rho(x,t)$ and $\int pFdp=p(x,t)\rho(x,t)$ one obtains
\bq\label{6.14}
\pp_t\rho+\pp_x\left[\frac{p(x,t)}{m}\rho(x,t)\right]=0
\end{equation}
Then multiply the Liouville equation by $p$ and integrate in order to obtain $({\bf 6E})\,\,
\int p^2F(x,p,t)dp=M_2(x,t)$ with
\bq\label{6.15}
\pp_t[\rho(x,t)p(x,t)]+\frac{1}{m}\pp_xM_2+(\pp_xV)\rho(x,t)=0
\end{equation}
Putting (6.14) into (6.15) one gets after some calculation
\bq\label{6.16}
\frac{1}{m}\pp_x[M_2(x,t)-p^2(x,t)\rho(x,t)]+\rho(x,t)\left[\pp_tp(x,t)+\pp_x\left(
\frac{p^2(x,t)}{2m}\right)+V_x\right]=0
\end{equation}
One can write then
\bq\label{6.17}
M_2-p^2(x,t)\rho(x,t)=\int[p^2-p^2(x,t)]F(x,p,t)dp=\int[p-p(x,t)]^2F(x,p,t)dp
\end{equation}
and set $({\bf 6F})\,\,\overline{(\gd p)^2}\rho(x,t)=\int[p-p(x,t)]^2F(x,p,t)dp$.  Then one
wants to find a functional expression for $\overline{(\gd p)^2}$ and we note here that
this expression is related to the momentum fluctuations used in \cite{c1,h2,h3,r1}.
Thus consider the entropy of an isolated system in the form $({\bf 6G})\,\,
{\mf S}(x,t)=k_Blog(\gO(x,t))$ where $k_B$ is the Boltzman constant and $\gO(x,t)$
represents the system accessible states when the position $x$ varies between $x$
and $x+\gd x$.  The equal a priori probability postulate then says that $({\bf 6H})\,\,
\rho(x,t)\propto \gO(x,t)=exp({\mf S}(x,t)/k_B)$ (cf. \cite{r2}).  One can now consider
entropy defined on configuration space and write $({\bf 6I})\,\,{\mf S}={\mf S}_{eq}+(1/2)
(\pp^2{\mf S}_{eq}/\pp x^2)(\gd x)^2$ where ${\mf S}_{eq}(x)$ is the statistical equilibrium
configuration entropy; here one has used the fact that the entropy must be a maximum
giving $(\pp {\mf S}_{eq}/\pp x)=0$ for $\gd x=0$.  One could also have divided the system
into N cells of dimension $\gd x$ and written (each $i$ refers to one specific cell)
\bq\label{6.18}
{\mf S}_i=({\mf S}_{eq})_i+(\pp_x{\mf S}_{eq})\gd x_i+\frac{1}{2}\left(\frac{\pp^2{\mf S}_{eq}}{\pp x^2}\right)
(\gd x_i)^2;
\end{equation}
Using known properties of entropy one can write ($\sum_1^N\gd x_i=0$ and the
system is adiabatically isolated)
\bq\label{6.19}
{\mf S}=\sum_1^N{\mf S}_i={\mf S}_{eq}+\frac{1}{2}\left(\frac{\pp^2{\mf S}}{\pp x^2}\right)\sum_1^N(\gd x_i)^2
\end{equation}
Now use ({\bf 6I}) to get
\bq\label{6.20}
\tl{\rho}(x,\gd x,t)=\rho_{eq}(x,t)exp\left(-\frac{1}{2k_B}\left|\frac{\pp^2{\mf S}_{eq}(x,t)}{\pp x^2}\right|(\gd x)^2\right)
\end{equation}
(recall that the second derivative of the entropy is negative
near an equilibrium point).
\\[3mm]\indent
Thus at each point $x$ the probability distribution with respect to small displacements
$\gd x(x,t)$ is Gaussian and is related with the probability of having a fluctuation
$\gD \rho$ owing to a fluctuation $\gd x$.  (6.20) guarantees that the system will tend to return to its equilibrium distribution represented by $\rho_{eq}$ and using (6.20)
the mean quadratic displacements related with the fluctuations are given via
\bq\label{6.21}
\overline{(\gd x)^2}=\frac{\int_{-\infty}^{\infty}(\gd x)^2exp(-\gag(\gd x)^2)d(\gd x)}
{\int_{-\infty}^{\infty}exp(-\gag(\gd x)^2d(\gd x)}=\frac{1}{2\gag}
\end{equation}
where $({\bf 6J})\,\,(1/2\gag)=k_B|\pp^2{\mf S}_{eq}(x,t)/\pp x^2|^{-1}$.  Note here
that $x$ is a constant so this is correct; it is $\gd x$ which is variable in the integration.
A priori there is no relation between the displacement and momentum fluctuations but
in the statistical equilibrium situation one can impose the restriction that 
$({\bf 6K})\,\,\overline{(\gd p)^2}\cdot\overline{(\gd x)^2}=\hbar^2/4$ (compare
here with the exact uncertainty principle of Hall and Reginatto in \cite{c1,h2,h3,r1}).
Using (6.21) and ({\bf 6K}) gives then
\bq\label{6.22}
\overline{(\gd p)^2}=-\frac{\hbar^2}{4}\frac{\pp^2log(\rho(x,t))}{\pp x^2}\Rightarrow
\overline{(\gd p)^2}\rho(x,t)=-\frac{\hbar^2}{4}\rho(x,t)\frac{\pp^2log(\rho(x,t))}{\pp x^2}
\end{equation}
Putting this in (6.16) and setting $\rho=R^2$ and $p=S_x$ one arrives at
\bq\label{6.23}
R^2\pp_x\left[S_t+\frac{1}{2m}S_x^2+V+Q\right]=0
\end{equation}
which together with (6.14) is equivalent to the SE as before, namely
\bq\label{6.24}
-\frac{\hbar^2}{2m}\psi_{xx}+V\psi=i\hbar\psi_t;\,\,\psi=Rexp(iS/\hbar)
\end{equation}
Thus (6.14) and (6.16) have the same content as the SE and one notes that
$p\sim S_x$ is also an assumption (pilot wave condition). 
\\[3mm]\indent
To connect this with $Z_Q$ and the previous derivation of the SE we recall (6.1) and
note that $\rho(x,t)=lim_{\gd x\to 0}Z_Q(x,\gd x,t)$ with $({\bf 6L})\,\,
\int p^2F(x,p,t)dp=lim_{\gd x\to 0}[-\hbar^2(\pp^2Z_Q(x,\gd x,t)/\pp(\gd x)^2)]$.
Then the right side of (6.16) can be written as
\bq\label{6.25}
\int[p-p(x,t)]^2F(x,p,t)dp=\lim_{\gd x\to 0}\left[-\hbar^2\frac{\pp^2Z_Q(x,\gd x,t)}
{\pp(\gd x)^2}+\frac{\hbar^2}{Z_Q}\left(\frac{\pp Z_Q}{\pp(\gd x}\right)^2\right]
\end{equation}
This can be rearranged as
\bq\label{6.26}
\overline{(\gd p)^2}\rho(x,t)=-\hbar^2lim_{\gd x\to 0}Z_Q(x,\gd x,t)\frac{\pp^2log(Z_Q)}
{\pp(\gd x)^2}
\end{equation}
It remains now to give the explicit appearance of this expression and show that it is equivalent to (6.22).  Note that $Z_Q$ can be written as
\bq\label{6.27}
Z_Q=\int F(x,p,t)dp+\frac{i\gd x}{\hbar}\int pFdp-\frac{(\gd x)^2}{2\hbar^2}\int p^2Fdp
+o((\gd x)^3)
\end{equation}
and this is equivalent (using ({\bf 6D}) and ({\bf 6E})) to
\bq\label{6.28}
Z_Q=\rho(x,t)+\frac{i\gd x}{\hbar}p(x,t)\rho(x,t)-\frac{(\gd x)^2}{2\hbar^2}M_2(x,t)
+o((\gd x)^3)
\end{equation}
The left side has to be written as $Z_Q=\psi^*(x-(\gd x/2))\psi
(x+(\gd x/2))$ and using $\psi=Rexp(iS/\hbar)$ one finds (up to second order in
the infinitesimal parameter)
\bq\label{6.29}
Z_Q=\left\{R^2+\left(\frac{\gd x}{2}\right)^2\left[RR_{xx}-(R_x)^2\right]\right\}
exp\left(\frac{i}{\hbar}S_x\gd x\right)
\end{equation}
Explicitly this is
\bq\label{6.30}
Z_Q=R^2+\frac{i\gd x}{\hbar}R^2S_x+\frac{(\gd x)^2}{2}\left[\frac{1}{4}R^2\frac
{\pp^2log(R^2}{\pp x^2}-\frac{R^2}{\hbar^2}(S_x)^2\right]
\end{equation}
Comparison with (6.28) gives $({\bf 6M})\,\,\rho(x,t)=R^2(x,t)$ and $p(x,t)=S_x(x,t)$  and $({\bf 6N})\,\,M_2(x,t)=-(\hbar^2/4)\rho(x,t)\pp^2log(\rho
(x,t))+p(x,t)^2\rho(x,t)$.  Using (6.28) one can write then
\bq\label{6.31}
Z_Q=\rho(x,t)\left[1+\frac{i\gd x}{\hbar}p(x,t)+\frac{(\gd x)^2}{2}\left[\frac{1}{4}
\pp^2_xlog(\rho(x,t))-\frac{p^2(x,t)}{\hbar^2}\right)\right]
\end{equation}
which means that
\bq\label{6.32}
lim_{\gd x\to 0}\frac{\pp^2}{\pp(\gd x)^2}log(Z_Q(x,\gd x,t)=\frac{1}{4}\pp^2_xlog(\rho(x,t))
\end{equation}
where one has expanded the logarithm in (6.31) up to the second order in $\gd x$.
This last result gives then, using (6.26),
\bq\label{6.33}
\overline{(\gd p)^2}\rho(x,t)=-\frac{\hbar^2}{4}\rho(x,t)\pp^2_xlog(\rho(x,t))
\end{equation}
which is equivalent to (6.22) as desired (this comes directly from ({\bf 6N})
and (6.17)).  Another way of comparing the two approaches is to simply substitute
(6.28) in the equation satisfied by the characteristic function which is (via (6.3))
$({\bf 6O})\,\,-i\hbar\pp_tZ_Q-(\hbar^2/m)(\pp^2Z_Q/\pp x\pp(\gd x))+\gd x(V_x)Z_Q=0$.
Taking the real and imaginary parts gives then
\bq\label{6.34}
\pp_t\rho(x,t)+\pp_x\left[\frac{p(x,t)}{m}\rho(x,t)\right]=0;
\end{equation}
$$\pp_t[\rho(x,t)p(x,t)]
+\frac{1}{m}\pp_xM_2(x,t)+V_x\rho(x,t)=0$$
which are (6.14) and (6.15).  Thus the restriction ({\bf 6K}) is equivalent to postulating
the adequacy of the Wigner-Moyal infinitesimal transformation together with the restriction of writing $Z_Q=\psi^*(x-(\gd x/2),t)\psi(x+(\gd x/2),t)$.

\section{THE UNCERTAINTY PRINCIPLE}
\renewcommand{\theequation}{7.\arabic{equation}}
\setcounter{equation}{0}

We go now to \cite{b27,c1,c28,g16,h2,h3,h5,l1,n4,p2,r1,r3,s26,s25} for some results
involving fluctuations and the uncertainty principle.  In particular \cite{b27} exhibits
some fascinating relations between the Heisenberg uncertainty principle and the exact
uncertainty principle of Hall-Reginatto and we sketch some of this here.  The title of \cite{b27}
is a question ``Is QM based on an uncertainty principle" and we indicate how this is answered in \cite{b27} in terms of the quantum potential (QP).  One modifies the classical
mechanical definition of momentum uncertainty in order to satisfy certain transformation
rules.  This involves adding a new term to the classical quadratic momentum uncertainty which has to be proportional to the inverse of a measure of the quadratic position
uncertainty.  Then one imposes the Hall-Reginatto conditions of causality and additivity
of kinetic energy which leads to a complete specification of the functional dependence
of the new term requiring it to be essentially the QP (often associated with the
idea of quantization - modulo some arguments at times).  An observer is now characterized by parameters denoting the statistical position and momentum uncertainties
$\gD x$ and $\gD p$ of its instruments.  For example $\gD x^2$ could be the trace of the covariance matrix associated to a given position probability density $\rho(x)$ or as the
inverse of the trace of the Fisher matrix associated to $\rho$.  The main postulate here is
that under dilatations of space coordinates the parameters $\gD x$ and $\gD p$ must
transform in such a manner that the relation $({\bf 7A})\,\,\gD x\gD p\geq (\hbar/2$ is
kept invariant.  The transformations allowed here are
\bq\label{7.1}
\gD x^{'2}=e^{-\ga}\gD x^2;\,\,\gD p^{'2}= e^{-\ga}\gD p^2+\frac{\hbar^2}{4}(e^{\ga}-
e^{-\ga})\frac{1}{\gD x^2}
\end{equation}
where $\ga$ is real and one sees that such transformations generate a group.
Multiplying the terms together in (7.1) gives
\bq\label{7.2}
\gD x^{'2}\gD p^{'2}=e^{-2\ga}\gD x^2\gD p^2+\frac{\hbar^2}{4}(1-e^{-2\ga})
\end{equation}
One notes that for $\ga\to \infty$ one has $\gD x^{'2}\gD p^{'2}\to (\hbar^2/4)$,
if $\gD x^2\gD p^2=\hbar^2/4$ then it remains so, and for $\ga\to -\infty$ one has
$\gD x^{'2}\gD p^{'2}\to\infty$ for any value of $\gD x^2\gD p^2\geq(\hbar^2/4)$.
Uniqueness of such transformations has not been established.  There is some analogy
here to special relativity and this is discussed in \cite{b27}.  In any case one shows
now that the stipulations (7.1) impose a radical modification of the laws of dynamics
that corresponds precisely to that required in the passage from classical to quantum
mechanics.  Thus consider $\rho(x)$ and $S(x)$ as basic variables (fields) and specify
that the time evolution of any functional via $({\bf 7B})\,\,{\mf A}=\int d^3xF(x,\rho,\na\rho,
\na^2\rho,\cdots,S,\na S,\na^2S,\cdots)$ is given by $({\bf 7C})\,\,\pp_t{\mf A}=
\{{\mf A},{\mf H}_{cl}\}$ where ${\mf H}_{cl}=\int d^3x(\rho|\na S|^2/2m)$ and
\bq\label{7.3}
\{{\mf A},{\mf B}\}=\int d^3x\left[\frac{\gd {\mf A}}{\gd \rho(x)}\frac{\gd {\mf B}}{\gd S(x)}
-\frac{\gd {\mf B}}{\gd \rho(x)}\frac{\gd {\mf A}}{\gd S(x)}\right]
\end{equation}
The Poisson bracket provides a Lie algebra structure.  Applying ({\bf 7C}) to $\rho$ and
$S$ yields
\bq\label{7.4}
\pp_t\rho=-\na\cdot\left(\rho\frac{\na S}{m}\right);\,\,\pp_tS=-\frac{|\na S|^2}{2m}
\end{equation}
(note $\na S\sim p$ which here is the classical momentum).  Now consider the group
of space dilatations $x\to exp(-\ga/2)x$ and its action on $\rho$ and $S$ via $({\bf 7D})\,\,
\rho'(x)=exp(3\ga/2)\rho(exp(\ga/2)x)$ and $S'(x)=exp(-\ga)S(exp(\ga/2)x)$ ($\ga$ real).
Such transformations preserve the normalization of $\rho$ and keep the dynamical
equations (7.3) invariant.  Assume now that the average momentum of the particle
is vanishing (i.e. a comoving frame is chosen which does not reduce the generality
of the results).  Then the classical definition of the scalar quadratic momentum 
uncertainty is given via $({\bf 7E})\,\,\gD p^2_{cl}=\int d^3x\rho|\na S|^2=2m{\mf H}_{cl}$
and under ({\bf 7D}) this becomes $({\bf 7F})\,\,\gD\rho^{'2}_{cl}=exp(-\ga)\gD p^2_{cl}$.  Also
any definition of the scalar quadratic position uncertainty measuring the dispersion 
$\gD x^2$ of
$\rho(x)$ transforms as $({\bf 7G})\,\,\gD x^{'2}=exp(-\ga)\gD x^2$.
The requirement of (7.1) to be fundamental now requires one to modify ({\bf 7E}) in order
to get a quantity whose variance satisfies (7.1).  Some argument shows that adding
a supplementary term proportional to $\hbar^2$ is needed in order to have $\gD p^2$
transform reasonably under ({\bf 7D}).   This is accomplished via
\bq\label{7.5}
\gD p_q^2=\int d^3x\rho(x)|\na S(x)|^2 +\hbar^2{\mf Q}
\end{equation}
where ${\mf Q}$ is to be determined.  Applying ({\bf 7D}) to (7.4) leads to $({\bf 7H})\,\,
\gD p^{'2}_q=e^{-\ga}\gD p_{cl}^2 +\hbar^2({\mf Q}'-e^{-\ga}{\mf Q})$. 
Adding and subtracting $exp(\ga)\hbar^2{\mf Q}$ yields and equation $({\bf 7I})\,\,\gD p^{'2}_q=
exp(-\ga)\gD p_q^2+\hbar^2({\mf Q}'-exp(-\ga){\mf Q})$.  Identifying this with (7.1) requires
$({\bf 7J})\,\,{\mf Q}'-exp(-\ga){\mf Q}=(1/4\gD x^2)(exp(\ga)-exp(-\ga))$ which, using (7.1)
becomes $({\bf 7K})\,\,{\mf Q}'-(1/4\gD x^{'2})=exp(-\ga)[{\mf Q}-(1/4\gD x^2)]$.  There are
an infinity of solutions but the form indicates a relation between ${\mf Q}$ and $\gD x^2$ that
is scale independent, namely $({\bf 7L})\,\,{\mf Q}=(1/4\gD x^2)$.  This is the only solution
for which the relation between $\gD p_q^2$ and $\gD x^2$ is independent of the scale $\ga$.
Thus this is the needed term to obtain a definition of $\gD p_q^2$ compatible with (7.1).
Since $\gD x^2$ depends only on $\rho(x)$ we see that ${\mf Q}$ is a functional of the form
({\bf 7B}) that does not depend on $S$.  For the precise form of $\gD x^2$ one can now refer to the Hall-Reginatto results (cf. \cite{c1} for a survey and a discussion
of relations to Fisher information)
which work with entirely
different arguments in discovering the additive term $\hbar^2{\mf Q}$.
At this point their additional
requirements of causality and additivity can be invoked point to obtain ${\mf H}_q=\gD p_q^2/2m$ and ${\mf Q}=\gb\int d^3x|\na\rho(x)^{1/2}|^2$
(with $\gb =1$) and finally
\bq\label{7.6}
{\mf H}_q=\int d^3x\left[\frac{\rho(x)|\na S|^2}{2m}+\frac{\hbar^2}{2m}|\na\rho(x)^{1/2}|^2\right]
\end{equation}
The HJ equation $\pp_tS=-(|\na S|^2/2m)+(\hbar^2/2m)(\na^2\rho(x)^{1/2}/
\rho^{1/2}(x)$ is given via ({\bf 7C})
along with the continuity equation in (7.3) leading to the standard SE for
$\psi=\rho^{1/2}exp(iS/\hbar)$.
\\[3mm]\indent
One looks next at dilatations under ({\bf 7D}) to get
\bq\label{7.7}
{\mf H}'_q=Cosh(\ga)\int d^3x\left[\frac{\rho|\na S|^2}{2m}+\frac{\hbar^2}{2m}|\na\rho^{1/2}|^2
\right]-
\end{equation}
$$-Sinh(\ga)\int d^3x\left[\frac{\rho|\na S|^2}{2m}-\frac{\hbar^2}{2m}|\na\rho^{1/2}|^2\right]$$
where $({\bf 7M})\,\,{\mf H}'_q[\rho,S]={\mf H}_q[\rho',S']$.  One writes then
\bq\label{7.8}
{\mf K}_q=\int d^3x\left[\frac{\rho|\na S|^2}{2m}-\frac{\hbar^2}{2m}|\na\rho^{1/2}|^2\right]
\end{equation}
In more compact notation now write $({\bf 7N})\,\,{\mf H}_q'=Cosh(\ga){\mf H}_q-Sinh(\ga)
{\mf K}_q$ with ${\mf K}_q'=-Sinh(\ga){\mf H}_q+Cosh(\ga){\mf K}_q$ so that under ({\bf 7D})
$({\mf H}_q,{\mf K}_q)$ transforms as a 2-D Minkowski vector under a Lorentz like transformation.  This corresponds to $({\bf 7O})\,\,t'=Cosh(\ga)t+Sinh(\ga)\tau$ and $\tau'=
Sinh(\ga)t+Cosh(\ga)\tau$ and one can write $({\bf 7P})\,\,\pp_t{\mf A}=\{{\mf A},{\mf H}_q\}$
with $\pp_{\tau}{\mf A}=\{{\mf A},{\mf K}_q\}$; further this transforms via ({\bf 7D}) to
$({\bf 7Q})\,\,\pp_{t'}{\mf A}'=\{{\mf A}',{\mf H}'_q\}$ and $\pp_{\tau'}{\mf A}'=\{{\mf A}',{\mf K}_q'\}$.
Thus this is all covariant under ({\bf 7D}).  The SE is a particular case of ({\bf 7P}) where
\bq\label{7.9}
{\mf A}=\psi=\rho^{1/2}e^{iS/\hbar};\,\,i\hbar\pp_t\psi=-\frac{\hbar^2}{2m}\na^2\psi
\end{equation}
and one obtains also
\bq\label{7.10}
i\hbar\pp_{\tau}\psi=-\frac{\hbar^2}{2m}\na^2\psi+\frac{\hbar^2}{m}\frac{\na^2|\psi|}{|\psi|}
\end{equation}
This is an interesting equation to be discussed later (see e.g. \cite{a2,d2,g30,g31,h30,s30,v1,w30}).
Thus (7.8) and (7.9) are covariant under space dilatations with
\bq\label{7.11}
i\hbar\pp_{t'}\psi'=-\frac{\hbar^2}{2m}\na^2\psi';\,\,i\hbar\pp_{\tau'}\psi'=-\frac{\hbar^2}{2m}
\na^2\psi'+\frac{\hbar^2}{m}\psi'\frac{\na^2|\psi'|}{|\psi'|}
\end{equation}
where (cf. ({\bf 7D}))
\bq\label{7.12}
\psi'=e^{3\ga/4}[\psi(e^{\ga/2}x)]^{(1/2)(1+exp(-\ga))}[\psi^*(e^{\ga/2}x)]^{(1/2)(1-exp(-\ga))}
\end{equation}
Note that when the transformation of the SE under dilatations is considered $\psi$ transforms
as the square root of a density $\rho^{1/2}$ (cf. also \cite{b3,b5,g1,g14,g12,h4}).
Finally one notes that defining ${\mf s}=\int d^3x\rho(x)S(x)$ (ensemble average of the quantum
phase up to a factor of $\hbar$) one has
\bq\label{7.13}
\{{\mf s},{\mf H}_q\}=\int d^3x\left[\frac{\rho|\na S|^2|}{2m}-\frac{\hbar^2}{2m}|\na\rho^{1/2}|^2
\right]={\mf K}_q
\end{equation}
$$\{{\mf s},{\mf K}_q\}=\int d^3x\left[\frac{\rho|\na S|^2}{2m}+\frac{\hbar^2}{2m}|\na\rho^{1/2}|^2
\right]={\mf H}_q$$
This leads to a general 
statement ${\mf A}'={\mf A}+\gd\ga\{{\mf A},{\mf s}\}$
\\[3mm]\indent
{\bf REMARK 7.1.}
Let us gather together some of the ideas connecting \cite{b27,h2,h3,h5,r1}.  Thus
in \cite{b27} (for 1-D here)
\bq\label{7.14}
\gD x^2\sim\frac{1}{\int[(\rho')2/\rho]dx}=\frac{1}{F}
\end{equation}
and the particle mean square deviation (or variance) is $\gs_x^2$ with $\gs_x^2\geq\gD x^2$.  To check this one can follow \cite{f1} and write for suitable estimators $\hat{\gt}$ of $\gt$ (note $\rho_{\gt}=\pp_{\gt}\rho=\rho \pp_{\gt} log(\rho)$ and $\hat{\gt}=\hat{\gt}(y)$)
\bq\label{7.15}
0=<\hat{\gt}-\gt>=\int dy\rho(y|\gt)[\hat{\gt}(y)-\gt]\Rightarrow 0=\int dy\rho_{\gt}
[\hat{\gt}-\gt]-\int\rho dy\Rightarrow
\end{equation}
$$\Rightarrow 1=\int dy\rho_{\gt}[\hat{\gt}-\gt]=\int dy\sqrt{\rho}[\hat{\gt}-\gt]\sqrt{\rho}\pp_{\gt}log(\rho)\leq \int \rho|\hat{\gt}-\gt|^2\int\rho|\pp_{\gt}log(\rho)|^2$$
This says that $1\leq \gs_x^2F$ with $\gs_x^2$ the classical variance and $F\sim$
Fisher information.  Now recall from $\#5$ before Example 5.1 that it is natural
to think of momentum fluctuations in the form $\gd p\sim\rho'/\rho=\pp_xlog(\rho)
\sim cu$ where $u$ is an osmotic velocity.  Then look at the exact uncertainty 
principle of \cite{h2,h3,h5} where one writes $\gd x\gD p_{nc}=\hbar/2$ (recall
$p_{nc}\sim \na S+\gd p$ with $\gd p\sim \hat{c}\pp log(\rho)$.  In \cite{o1}
on the other hand there is a crucial condition $({\bf 6K})\,\,\overline{(\gd p)^2}
\overline{(\gd x)^2}=\hbar^2/4$ which is imposed in the context of a statistical
equilibrium situation involving Boltzman type entropy.  Here $\overline{(\gd p)^2}$
is determined via (({\bf 6K}) and (6.21) in the form $({\bf 7R})\,\,\overline{(\gd x)^2}
=1/2\gag$ where $2\gag=(1/k_B)\pp^2{\mf S}_{eq}/\pp x^2$ leading to (6.22)
where $({\bf 7S})\,\,\overline{(\gd p)^2}\sim-(\hbar^2/4)\pp_x^2log(\rho)$.  Since
$\pp^2log(\rho)\sim\pp(\rho'/\rho)=(\rho''/\rho-(\rho'/\rho)^2$ one sees that for 
$\rho'=\rho=0$ outside of a compact $\gO$ $({\bf 7T})\,\,\int \rho\pp^2log(\rho)=
-\int(\rho')^2/\rho=-\int\rho(\rho'/\rho)^2$ and such quantum perturbations 
to $\na S)^2$ would involve the same action as perturbations $(\rho'/\rho)^2$
(corresponding to $\int \rho Q$ where $Q\sim (\pp^2\sqrt{\rho})/\sqrt{\rho}=
(1/2)[(\rho''/\rho)-(\rho^{'2}/2\rho^2]$.$\hfill\bs$
\\[3mm]\indent
We go next to \cite{s26} and sketch some of the results.  In ``geometric quantum
mechanics" (GQM) geometry is not prescribed but is determined by the space matter content.
Further in \cite{s26} one assumes that the affine connections are responsible for
quantum phenomena.  Given that GQM deals with a Gibbs ensemble of particles,
rather than a single particle, one can treat it as a classical theory.  Wave equations and
the operational prescriptions of standard QM are devoid of physical meaning and 
are regarded as clever devices to overcome the difficulties inherent in a nontrivial
geometric structure (see \cite{c1,s1} and Section 2 for more on this).  The object in
\cite{s26} is to show how the Heisenberg uncertainty principle arises in the context of GQM.
Note that Planck's constant is conspicuously absent in the treatment.
Thus in GQM one assumes that in a parallel displacement $d{\bf r}$ the length 
$\ell=({\bf A}\cdot{\bf A})^{1/2}$ of a vector {\bf A} changes by an amount 
$(\bullet)\,\,\gd\ell=\ell(\vec{\phi}\cdot d{\bf r})$ where $\vec{\phi}$ is the Weyl gauge
vector.
In nonrelativistic mechanics the spatial metric is Euclidean so the dot means the standard
scalar product and we have a Weyl space which generally will have a nonzero scalar
curvature ${\mc R}$ even if its metric has zero curvature.  In GQM the law $(\bullet)$
and hence the Weyl-Ricci curvature is determined by the motion of the particle itself and
in turn it acts as a guidance field for the particle (this is perhaps another way to look at Bohmian
mechanics).  Following \cite{s1} (cf. also Section 2) the length transference law
may be obtained from a minimum average curvature principle $(\bullet\bullet)\,\,
E[{\mc R}({\bf r}(t),t)]=min$ where $E$ denotes the ensemble expectation value.  When
the space is flat ($\dot{R}=0$) one has from (2.3) (cf. \cite{c1,s1}) $(\bl)\,\,{\mc R}=
2|\vec{\phi}|^2-4\na\cdot\vec{\phi}$.  Now let $\rho$ be the probability density of the particle
position (properly normalized and vanishing at infinity) and note that
\bq\label{7.16}
E[\na\cdot\vec{\phi}]=\int_{{\bf R}^3}\rho\na\cdot\vec{\phi}d^3{\bf r}=-\int_{{\bf R}^3}
\rho\vec{\phi}\cdot\na(log(\rho)d^3{\bf r}=-E[\vec{\phi}\cdot\na log(\rho)]
\end{equation}
Consequently
\bq\label{7.17}
E[{\mc R}]=2E[|\vec{\phi}|^2+2\vec{\phi}\cdot\na log(\rho)]=2E[(\vec{\phi}+\na log(\rho)^2]-
2E[(\na log(\phi)^2]
\end{equation}
The minimizing gauge vector $\phi$ is found from this to be $(\bl\bl)\,\,\vec{\phi}=
-\na log(\rho)$ and the minimum average scalar curvature is $(\bgs)\,\,E[{\mc R}]=
-2E[|\vec{\phi}|^2]$.  Note that the minimal average curvature is negative and moreover from
$(\bl\bl)$ one obtains
\bq\label{7.18}
E[\vec{\phi}({\bf r},t),t)]=\int_{{\bf R}^3}\rho({\bf r},t)\vec{\phi}({\bf r},t)d^3{\bf r}=
-\int_{{\bf R}^3}\rho\na log(\rho)d^3{\bf r}=-\oint_S\rho{\bf n}dS=0
\end{equation}
where S is a closed surface at $\infty$ enclosing ${\bf R}^3$ (where $\rho=0$).  One notes
also that the mean square deviation for $|\vec{\phi}|$ is not zero in general.  In fact from 
$(\bgs)$ and (7.18) one finds the mean square deviation via
\bq\label{7.19}
\gD|\vec{\phi}|=\{E[|\vec{\phi}|^2]-E^2[\vec{\phi}]\}^{1/2}]=2^{-1/2}[E(-{\mc R}]^{1/2}
\end{equation}
Now let $\gD q$ be the root mean square deviation of the particle position so that
(7.18) and the Schwarz inequality give
\bq\label{7.20}
\gD q\gD |\vec{\phi}|\geq Cov\{({\bf r}-E[{\bf r}])\cdot (\vec{\phi}-E[\vec{\phi}])\}=
E[{\bf r}\cdot\vec{\phi}]
\end{equation}
where Cov denotes covariance.  The average on the right side of (7.20) can be computed
using $(\bl\bl)$ and one gets (using (7.19))
\bq\label{7.21}
\gD q\gD |\vec{\phi}|^2\geq \int_{{\bf R}^3}\rho\na\cdot{\bf r}d^3{\bf r}=3\Rightarrow
\gD q(E[-{\mc R}])^{1/2}\geq \frac{3}{\sqrt{2}}
\end{equation}
This is the fundamental relation between the particle root mean square deviation and the
average space curvature.  The space may be flat on average only if the particle is completely
delocalized - particle localization forces the space to be curved.  One notes that (7.21) is
purely geometrical and Planck's constant is not involved.  However (7.21) does imply
the Heisenberg uncertainty principle provided the prescriptions of standard QM are used.
Thus for $\psi=\sqrt{\rho}exp(iS/\hbar)$ with a particle of mass $m$ and probability density
$\rho=|\psi|^2$ one  has trajectories $(\bgs\bgs)\,\,d{\bf r}/dt=\na S({\bf r},t)/m$ where
\bq\label{7.22}
S_t+\frac{|\na S|^2}{2m}+V-\frac{\hbar^2}{16m}{\mc R}=0
\end{equation}
(cf. Section 2 and \cite{s1}).  The averages obtained using the operator methods of standard
QM do not coincide in general with the averages made on the classical ensemble.  Thus
using $<\,\,,\,\,>$ for QM and $E[\,\,]$ for ensemble averages one gets (using (7.22))
\bq\label{7.23}
<{\bf r}>=E[{\bf r}];\,\,<{\bf p}>=E[\na S];\,\,<\hat{p}^2>=E[|\na S|^2]+\frac{\hbar^2}{8}E[-{\mc R}]
\end{equation}
where $-i\hbar\na\sim \hat{p}$.  Note here also that defining $\gD p=[<{\bf p}-
<{\bf p}>^2]^{1/2}$ it follows immediately via (7.23) that
\bq\label{7.24}
(\gD p)^2=E[(\na S-E[|\na S|)^2]+\frac{\hbar^2}{8}E[-{\mc R}]\Rightarrow \gD p\geq
\frac{\hbar}{2\sqrt{2}}(E[-{\mc R}])^{1/2}
\end{equation}
leading to the Heisenberg type inequality $({\bf 7U})\,\,\gD p\gd Q\geq (3/2)\hbar$
via (7.21).
\\[3mm]\indent
{\bf REMARK 7.2.}
The relation $\gD x\gD p\geq (\hbar/2)$ can be traced in an analogous manner using 
$\gD x\gD \phi_x\geq E[x\phi_x]$ in place of (7.20).$[\hfill\bs$
\\[3mm]\indent
Finally one notes that in (7.24) there is a random motion part and a curvature part and it is
the curvature part that forces Heisenberg's position momentum uncertainty relation.
However this contribution is a mere consequence of the operator formulation of QM and,
unlike (7.21), it has no physical meaning in a physically consistent approach to GQM.
In fact, GQM being a classical theory, the particle momentum should be defined as $p=
\na S$ and the mean square deviation should be given by the first term on the right in (7.24).
\\[3mm]\indent
{\bf REMARK 7.3.}
In reading over the second paper in \cite{s25} one is struck by similarities to the approach of Nottale
(see \cite{c1,c20,n4}).  The QP term arises basically because the quantum paths are not
differentiable (\`a la Feynman they have fractal dimension 2) and as indicated in \cite{c1} this corresponds to expressing the QP in terms of an osmotic velocity ${\bf u}$ (cf.
$\#1$ before (5.2) in Section 5).  In the Nottale approach {\bf u} only appears because
of ``jagged" paths and in the presence of smooth (i.e. differentiable) paths the QP is
zero and there is no SE.  If there are nonsmooth paths then the fact that ${\bf u}=D
\na log(\rho)$ (with $\vec{\phi}=-\na log(\rho)$) suggests that ${\bf u}$ is a curvature
phenomena but that is misleading - it is really a diffusion phenomena and for D=0
{\bf u} will not appear.  In more detail, following Nottale, one can write
\bq\label{7.25}
\rho_t+div(\rho b_{+})=D\gD\rho;\,\,\rho_t+div(\rho b_{-})=-D\gD\rho
\end{equation} 
But $b_{\pm}$ represents e.g. right and left derivatives at a given point so smooth
paths give $b_{+}=b_{-}$ and hence ${\bf u}\sim (1/2)(b_{+}-b_{-})=0$ with $D=0$
necessarily via $D=-D$.
Hence smooth paths imply no Wiener process and no SE.  One can still
have curvature via a nonzero Weyl vector $\phi\sim -\na log(\rho)$.  The SE and 
QM arise via an assumption of diffusion and nonsmooth paths which is equivalent to
introducing a nonzero QP.  This does not answer the question ``Why QM?" but
does seem to provide evidence that ({\bf 7U}) is the fundamental GQM idea and
the Heisenberg inequality arises only after introducing diffusion with a particular
diffusion coefficient $\hbar/2m$ (cf. \cite{d2}), or introducing a Schr\"odinger wave function containing
$\hbar$, or perhaps after introducing all the miracle machinery of Hilbert space QM, etc.
This is all consistent with comments made in Remarks 2.1 and 2.2.  The factor $\hbar$ is
``gratuitous" and is introduced in QM either via the wave function idea $\psi=Rexp(iS/\hbar)$ or via a diffusion coefficient $D=\hbar/2m$.
$\hfill\bs$

\section{GEOMETRY AND QUANTUM MATTER}
\renewcommand{\theequation}{8.\arabic{equation}}
\setcounter{equation}{0}

We have seen already a number of striking relations between geometry and quantum matter some of which we repeat here via equations.  Thus
\begin{enumerate}
\item
From (2.9) we have
\bq\label{8.1}
Q\sim-\frac{\hbar^2}{16m}\left[\dot{{\mc R}}+2\left\{\phi_i\phi^i-\frac{2}{\sqrt{g}}
\pp_i(\sqrt{g}\phi^i)\right\}\right]
\end{equation}
\item
From (2.10) follows
\bq\label{8.2}
\phi_k\phi^k-2\pp_k\phi^k\sim -\left(\frac{|\na\rho|^2}{\rho^2}-\frac{2\gD\rho}{\rho}\right)
=4\frac{\gD\sqrt{\rho}}{\sqrt{\rho}}
\end{equation}
\item
For n=4 from (2.18) results
\bq\label{8.3}
{\mc R}=\dot{{\mc R}}-3\left[\frac{1}{2}g^{\mu\nu}\phi_{\mu}\phi_{\nu}+\frac{1}{\sqrt{-g}}
\pp_{\mu}\sqrt{-g}g^{\mu\nu}\phi_{\nu}\right]={\mc R}+{\mc R}_w
\end{equation}
\item
From (2.3) for n=3 we have
\bq\label{8.4}
{\mc R}=\dot{{\mc R}}+2\left[\phi_i\phi^i-2\left(\frac{1}{\sqrt{g}}\pp_i(\sqrt{g}\phi^i)
\right)\right]
\end{equation}
\item
Further one knows that
\bq\label{8.5}
{\mf M}^2=m^2e^{Q};\,\,Q=\frac{\hbar^2}{m^2c^2}\frac{\bx \sqrt{\rho}}{\sqrt{\rho}}
=\frac{\hbar^2}{m^2c^2}\frac{\na^2|\psi|}{|\psi|}
\end{equation}
\item
The following equations also relate the Weyl vector $\vec{\phi}$ and the
Dirac field $\gb$
\bq\label{8.6}
\gb\sim{\mf M};\,\,\gb_0\to\gb=\gb_0e^{-\Xi};\,\,\phi_{\mu}\to\phi_{\mu}+\pp_{\mu}
\Xi
\end{equation}
\item
Further $Q\sim -(m^2/2){\bf u}^2-
(\hbar/2)\pp{\bf u}$ (${\bf u}\sim$ osmotic velocity) with
\bq\label{8.7}
{\bf u}=D\na log(\rho)=\frac{\hbar}{m}\frac{\na\sqrt{\rho}}{\sqrt{\rho}};
\,\,\vec{\phi}=-\na log(\rho);
\end{equation}
\end{enumerate}
\indent
Thus in particular the quantum mass $m_Q\sim{\mf M}$ is given via Q, or via
$\rho$, or via $\vec{\phi}$ so $\vec{\phi}$ determines ${\mf M}$ and conversely
${\mc M}$ determines $\rho$ via Q.  Recall that $\gD R+\gb QR=0$ has unique
solutions in say $H_0^1(\gO)$ where $\gb=-(2m/\hbar^2)$.  In fact $\gD R+\gb QR
=\gl R$ has a unique solution for $\gl\ne\gS$ where $\gS$ is a countable set
$\gS\subset {\bf R}$ (cf. \cite{c8}) so if $0\ne \gS$ then Q determines R uniquely.
In any event $\rho$ itself determines ${\mf M}$ and $\vec{\phi}$ (as well as {\bf u}).
Since $\rho$, or $\vec{\phi}$, or $\gb$ determine the Weyl Dirac geometry (the 
Riemannian metric is assumed fixed here) one seems to have already an excellent
theory for quantum perturbations of a classical Riemannian geometry.  Is this not
an important desideratum of quantum gravity theory?  To ask for quantum perturbations
that maintain a Riemannian structure seems perhaps excessive (cf. \cite{c17} 
(although conformal equivalence is established via ${\mf M}^2/m^2$) and we
are not trying to answer the questions of basic structure (for this see e.g.
\cite{b30,b31,c30,m1,i3,r4,z1}).

\section{REMARKS ON MANY WORLDS}
\renewcommand{\theequation}{9.\arabic{equation}}
\setcounter{equation}{0}

We refer back to Remark 2.2, Remark 2.5, and Sections 3-5.  Consider a 3-D
Riemannian manifold M with metric $g_{ij}$ and Riemann curvature $\dot{{\mc R}}$.
Let $m$ be a mass and consider an ensemble of particles of mass $m$
distributed via a probability density P which generates a mass density $\rho=
mP(x,t)$.  Write $\hat{\rho}=\rho/\sqrt{g}$ and observe that a natural classical
background for quantum evolution in M is provided via a Weyl geometry on M
based on a Weyl vector $\vec{\phi}=-\na log(\hat{\rho})$.  Then the SE (2.6)
for $\psi=Rexp(iS/\hbar)$, namely
\bq\label{9.1} 
i\hbar\psi_t-\frac{\hbar^2}{2m}\frac{1}{\sqrt{g}}[\pp_i(\sqrt{g}g^{ik}\pp_k)]\psi
+[V-\frac{\hbar^2}{16m}\dot{{\mc R}}]\psi
\end{equation}
corresponds to classical evolution (cf. ({\bf 2G}), (2.4), and (2.5))
\bq\label{9.2}
\pp_tS+\frac{1}{2m}g^{ik}\pp_iS\pp_kS+V-\frac{\hbar^2}{2m}{\mc R}=0;
\end{equation}
$$\pp_t\hat{\rho}+\frac{1}{m\sqrt{g}}\pp_i(\sqrt{g}\pp_iS\hat{\rho})=0$$
where
\bq\label{9.3}
{\mc R}=\dot{{\mc R}}+\frac{8}{\sqrt{g\hat{\rho}}}\pp_i(\sqrt{g}g^{ik}\pp_k\sqrt{\rho})];\,\,Q\sim -\frac{\hbar^2}{16m}{\mc R}
\end{equation}
Here ${\mc R}={\mc R}_w$ is the Ricci-Weyl curvature.
\\[3mm]\indent
{\bf REMARK 9.1.}
Thus assume M is Euclidean with $\dot{{\mc R}}=0$ for simplicity so $g\sim 1$ and 
$\hat{\rho}=\rho$.  The Weyl vector is then $\vec{\phi}=-\na log(\rho)$ and 
\bq\label{9.4}
{\mc R}=\frac{8\gD\sqrt{\rho}}{\sqrt{\rho}};\,\,Q\sim -\frac{\hbar^2}{2m}\frac{\gD
\sqrt{\rho}}{\sqrt{\rho}}
\end{equation}
This suggests a kind of many worlds scenario for quantum motion.  Thus given a
probability distribution $P(x,t)$ with $\rho=mP$ as above one has wave functions
$\psi=Rexp(iS/\hbar)$ where $R\sim\sqrt{\rho}$ (and some normalization
$\int R^2dx=\int \rho dx=1$ for example is imposed).  This seems to say that each such
wave function $Rexp(iS/\hbar)$ generates a family of time dependent Weyl geometries evolving via (9.2).  The choice of $R\sim\sqrt{\rho}$ determines the
geometry and we have an infinite number of Weyl space scenarios, each created by a choice of probability distribution $P(x,t)$.  Thus a quantum particle is associated to an infinite number of time dependent Weyl space paths, each determined by some particular wave function
$\psi=Rexp(iS/\hbar)$ where $R=|\psi|$ can be prescribed via a suitably normalized
probability distribution $P(x,t)$.  In view of (9.4) each $\rho$ as above is associated
to a quantum potential Q and if the association $R\leftrightarrow Q$ holds 
(as in Remark 5.1) then it could be said that Q determines the Weyl spaces.
If $P(x,t)=P(x)$ is independent of time the Weyl space is fixed.
$\hfill\bs$
\\[3mm]\indent
{\bf REMARK 9.2.}
One can rephrase the above as follows.  Take M as in Remark 9.1 with $\dot
{{\mc R}}=0$ and posit a SE (9.1) in the form $({\bf 9A})\,\,i\hbar\psi_t=-
(\hbar^2/2m)\gD\psi+V\psi$.  Every wave function $\psi=Rexp(iS/\hbar)$ gives rise
to a probability distribution $R^2=|\psi|^2=\rho$ and thence to a time dependent family of Weyl geometries
via $\vec{\phi}=\na log(\rho)$ within which $R$ and $S$ evolve classically via
(9.2) which we rewrite here as
\bq\label{9.5}
\pp_tS+\frac{1}{2m}S_x^2+V+Q=0;\,\,\pp_t\rho+\frac{1}{m}(\rho S_x)_x=0;\,\,
Q=-\frac{\hbar^2}{2m}\frac{\gD\sqrt{\rho}}{\sqrt{\rho}}
\end{equation}
Thus every wave function picks some worlds (families of Weyl spaces) describing
the evolution of R and S.
Recalling that $Q\sim -(\hbar^2/16m){\mc R}_w$ one could
also say that the quantum potential (defined via $\rho$) determines the family of Weyl
spaces and indicates a quantization of the classical evolution (9.5) (with $Q=0$).
Here $\hbar$ and $m$ arise via the SE and one could consider the original
situation $\rho=mP=R^2$ here.  If one expects $\int R^2dx=1$, which means
$m\int Pdx=1$ so that the choice of P is constrained via $\int Pdx=1/m$,
then it seems necessary to take $P=P(x)$ with a fixed Weyl space; in this situation
SE corresponds to an infinite number of
classical systems evolving in a fixed Weyl space, each determined by a wave function of the SE
with $S=S(\rho,V)$.$\hfill\bs$
\\[3mm]\indent
{\bf REMARK 9.3.}
It is not at all clear whether or not all this has any relation to the many worlds 
interpretation (MWI) (cf.  \cite{b10,b11,h10,s10,s11,t2,v2,w3,w4}) and we will
not belabor the idea.  In many respects Bohmian mechanics as well as the
consistent (or decoherent) histories approach (although different) seem to give a deeper insight into
quantum processes than does the standard Copenhagen framework.  Both are
based on trajectories but represent a given history by different operators.   We refer
here to \cite{b32,g3,g2,h7,h1,h8,h6} for more on this.
The important relation between quantum mechanics (QM) and mathematical
statistics elucidated in \cite{b12} will be discussed below and this is relevant to
any theory of wave functions.$\hfill\bs$
\\[3mm]\indent
Let us now examine the situation in Remarks 9.1 and 9.2 in a more coherent manner, based on the Ricci-Weyl curvature ${\mc R}={\mc R}_w\sim Q$ as the basic ingredient
(cf. (9.3) - (9.4)).  Thus let $\Xi$ be a class of quantum potentials, $Q=Q(x,t)$, for
which $\gD R+\gb QR=0\,\,(\gb>0)$ has a unique solutiion $R\in H_0^1(\gO)$
(where eventually $\gb\sim 2m/\hbar^2$).  The $t$ dependence in R arises directly from
Q in this context.  We specify now a Ricci-Weyl curvature $({\bf 9A)}\,\,{\mc R}_w
(x,t)=-(16m/\hbar^2)Q(x,t)=-8\gb Q(x,t)$ and think of $R(x,t)=R\sim
\sqrt{\rho}$.  We assume in the background a SE with potential V and via (9.5) one has
e.g.
\bq\label{9.6}
\frac{1}{m}(\rho S_x)(x,t)=-\int_0^x\pp_t\rho dx+F(t)
\end{equation}
where $(1/m)(\rho S_x)(0,t)=F(t)$ and $\rho(0,t)$ is known.  If we are dealing with
particle motion where $S_x\sim p$ then $S_x(0,t)\sim p(0,t)$ and we see also
that $S_t$ is known from (9.5) up to a factor $p(0,t)$.
Hence given ${\mc R}_w\sim -8\gb Q$ based on a SE with potential V we can determine
$S_x$ and $S_t$ up to a factor $p(0,t)=S_x(0,t)$ leading to
\begin{theorem}
Given a SE based on a potential V in a region $\gO\subset {\bf R}^3$ let ${\mc R}_w
\sim -8\gb Q$ be given with $Q\in \Xi$ ($\gb\sim 2m/\hbar^2$).  Then ${\mc R}_w$
can be associated with a wave function $\psi=Rexp(iS/\hbar)$ and a Weyl geometry
based on a Weyl vector $\vec{\phi}=-\na log(\rho)$ where $R=\sqrt{\rho}$ is completely
determined with $S_x$ and $S_t$ known up to a factor $S_x(0,t)\sim p(0,t)\sim F(t)$.
Thus for $Q\in \Xi$ the quantum potential, or equivalently the Weyl-Ricci curvature,
determines (modulo $F(t)$ and a constant) a Weyl space path describing the evolution of R and S with corresponding wave function $\psi$.
\end{theorem}
\indent
Suppose for a given ${\mc R}_w\sim -8\gb Q$ and its corresponding $R(x,t)$ one 
had different $S(x,t)$, e.g. $S$ and $\hat{S}$ with $S_x\sim F(t)$ and $\hat{S}_x\sim
\hat{F}(t)$ so that from (9.5) and (9.6)
\bq\label{9.7}
\frac{1}{m}\rho(\hat{S}_x\mp S_x)=\hat{F}\mp F=G_{\mp}(t)\Rightarrow
\hat{S}_x\mp S_x=\frac{mG_{\mp}(t)}{\rho(x,t)}
\end{equation}
\bq\label{9.8}
\hat{S}_t-S_t=-\frac{1}{2m}(\hat{S}_x^2-S_x^2)=-\frac{1}{2m}(\hat{S}_x-S_x)(\hat{S}_x
+S_x)=-\frac{m}{2\rho^2}(G_{-}G_{+})
\end{equation}
\begin{corollary}
Given a SE based on a potential V, a wave function $\psi$ determines a unique Weyl
space path and Weyl-Ricci curvatures ${\mc R}_w$ with corresponding evolution of R and S.
On the other hand a given Weyl-Ricci curvature ${\mc R}_w\sim -8\gb Q$ with 
$Q\in \Xi$ determines a unique Weyl space path for motion of R, S with R known completely and S determined modulo (9.7)-(9.8).
\end{corollary}
\begin{corollary}
As stated in Remark 9.2 a given SE is associated to a collection of Weyl space paths
(determined via wave functions) describing the evolution of R and S.  The Weyl space
path is completely characterized via $Q\in \Xi$ (i.e. by Weyl-Ricci curvatures).
\end{corollary}

\subsection{QUANTUM INFORMATICS}

There is a fascinating series of papers indicated in \cite{b12} and we only sketch here a few ideas.  Let $\psi(x)\in L^2({\bf R})$ be given and consider
\bq\label{9.9}
\psi(x)=\frac{1}{\sqrt{2\pi}}\int \tl{\psi}(p)e^{ipx}dp;\,\,\tl{\psi}(p)=\frac{1}{\sqrt{2\pi}}
\int \psi(x)e^{-ipx}dx
\end{equation}  
where (naturally) $\int|\psi|^2dx=\int |\tl{\psi}|^2dp$. 
One writes $({\bf 9D})\,\,P(x)=|\psi(x)|^2$
and $\tl{P}(p)=|\tl{\psi}(p)|^2$ with normalization 
$({\bf 9E})\,\,\int P(x)dx=\int\tl{P}(p)dp=1$.  
The coordinate and momentum probability distributions $P(x)$ and $\tl{P}(p)$ are called
mutually complementary statistical distributions and one notes that e.g. information about
the phase S of $\psi$ is lost.  For an experimental extracting of information it is not sufficient to use only one fixed representation (Bohr's complementarity principle).
There is no distribution $P(x,p)$ corresponding to the $P(x)$ and $\tl{P}(p)$ distributions
since this would violate the Heisenberg uncertainty principle.  One shows in \cite{b12}
that classical statistics is incomplete here while quantum statistics is complete
via the introduction of $\psi$. 
Thus
write
\bq\label{9.10}
P(x)=\psi^*\psi=\frac{1}{2\pi}\int dpdp_1\tl{\psi}^*(p)\tl{\psi}(p_1)e^{-ix(p-p_1)}=
\end{equation}
$$\frac{1}{2\pi}\int dudp\tl{\psi}^*(p)\tl{\psi}(p-u)e^{-ixu}=\frac{1}{2\pi}\int f(u)e^{-ixu}du$$
where $({\bf 9F})\,\,f(u)=\int dp\tl{\psi}^*(p)\tl{\psi}(p-u)=\int dp\tl{\psi}^*(p+u)\tl{\psi}(p)$.
Thus one has $({\bf 9G})\,\,P(x)=(1/2\pi)\int f(u)exp(-ixu)du$ with $({\bf 9H})\,\,
f(u)=M(exp(iux))=\int P(x)exp(ixu)dx$ (mean value).  Similarly $({\bf 9I})\,\,
\tl{f}(t)=\int \tl{P}(p)exp(ipt)dp=M(exp(ipt))$ for $({\bf 9J})\,\,\tl{f}(t)=\int dx\psi^*(x-t)
\psi(x)$.  This leads to the assertion that in order for the function $f(u)$ to be a characteristic function (as above) it is necessary and sufficient that it be represented
as a convolution as in ({\bf 9F}) with $\tl{\psi}$ satisfying $\int dp|\tl{\psi}(p)|^2=1$.
To see the necessity let $f(u)$ be a characteristic function so by ({\bf 9G}) it defines a 
density $P(x)$.  Let $\psi(x)=\sqrt{P(x)}exp(iS(x))$ for arbitrary real S (e.g. $S=0$); this
amounts to completing a classical statistical distribution to a quantum state.  Then
$\tl{\psi}(p)$ defined via (9.9) provides the decomposition ({\bf 9F}) (thus $f(u)\to\tl{\psi}
(p)$).  For sufficiency let $f(u)$ be represented via ({\bf 9F}) via $\tl{\psi}(p)$ (normalized
as in ({\bf 9D})).  Then $f(u)$ will be a characteristic function for $P(x)$ defined via ({\bf 9G}) (and normalized).  Thus $\tl{\psi}(p)\to f(u)\to P(x)$.  We note that even in classical
statistics the equation ({\bf 9F}) implicitly reveals the existence of a momentum space and the corresponding wave function $\tl{\psi}(p)$ and suggests the paucity of classical 
notions of probability.  Of course from ({\bf 9F}) one cannot derive a unique wave function
$\tl{\psi}(p)$ and ({\bf 9D}) does not yield $\psi(x)$ unambiguously.  Hence one classical
probability distribution may be described by a number of quantum objects; in order for the statistical theory to be ``complete" it has to be expanded in a manner to obtain a quantum
state vector $\psi$ (e.g. via introduction of a phase multiplier).
\\[3mm]\indent
One notes that moments of a random variable can be calculated via means of the 
characteristic function.  Thus for $f(u)=\int P(x)exp(ixu)dx$ one has 
$({\bf 9K})\,\,f^{(k)}(0)=i^kM(x^k)\,\,(k=0,1,2,\cdots)$.  Simple calculations then lead to
representations $\hat{x}\sim i\pp_p$ in momentum space and $\hat{p}=-i\pp_x$ in
coordinate space.  Consequently e.g. $\hat{p}\hat{x}-\hat{x}\hat{p}=-i$ is invariant 
under change of representation space.  We recall also the standard 
Cauchy-Schwartz-Bunyakowski inequality $({\bf 9L})\,\,|<\phi|\psi>|^2\leq <\phi|\phi>
<\psi|\psi>$ and one introduces a ``fidelity" F via $({\bf 9M})\,\,F=(|<\phi|\psi>|^2/
<\phi|\phi><\psi|\psi>)$ with $0\leq F\leq 1$.  The Heisenberg uncertainty relation can be
derived via consideration of 
\bq\label{9.11}
F(\xi)=<\psi|(-i\xi\hat{p}+\hat{x})(i\xi\hat{p}+\hat{x})\psi>=\xi^2M(\hat{p}^2)-i\xi M(\hat{p}
\hat{x}-\hat{x}\hat{p})+M(\hat{x}^2)\geq 0
\end{equation}
Setting e.g. $({\bf 9N})\,\,D_p=M(\hat{p}^2)-(M(\hat{p}))^2$ this leads to $({\bf 9O})\,\,
D_xD_p\geq 1/4$ with equality only when $(i\xi\hat{p}+\hat{x})|\psi>=0$ for some $\xi$
(which means Gaussian states).  A more general result is the Robertson-Schr\"odinger
uncertainty relation (cf. \cite{b12,r5}).  Thus let $z_1$ and $z_1$ be two observables
(centered so $(M(z_1)=M(z_2)=0$) and consider
\bq\label{9.12}
F(\xi)=<\psi|(\xi exp(-i\phi)z_2+z_1)(\xi exp(i\phi)z_2+z_1)|\psi>\geq 0
\end{equation}
Here $\xi$ and $\phi$ are real and one defines the covariance as $({\bf 9P})\,\,
cov(z_1,z_2)=(1/2)<\psi|z_1z_2+z_2z_1|\psi>$.  Let $z_1z_2-z_2z_2=iC$ where
C is Hermitian so that $({\bf 9Q})\,\,M(C)=-i<\psi|z_1z_2-z_2z_1|\psi>$ leading to
\bq\label{9.13}
F(\xi)=\xi^2M(z^2_2)+\xi[2cov(z_1,z_2)cos(\phi)-M(C)Sin(\phi)]+M(z_1^2)
\end{equation}
Set $\rho^2=4(cov(z_1,z_2))^2+(M(C))^2$ and choose an angle $\gb$ so that
\bq\label{9.14}
2cov(z_1,z_2)=\rho Cos(\gb)\,\,and\,\,M(C)=\rho Sin(\gb)\Rightarrow 
\end{equation}
$$\Rightarrow F(\xi)=
\xi^2M(z_2^2)+\xi\rho Cos(\phi+\gb)+M(z_1^2)\geq 0$$
Now choose $\phi$ so that $Cos(\phi+\gb)=1$ which yields 
\bq\label{9.15}
M(z_1^2)M(z_2^2)=D(z_1)D(z_2)\geq \frac{\rho^2}{4}=\left((cov(z_1,z_2))^2+
\frac{(M(C))^2}{4}\right)
\end{equation}
Then define a correlation coefficient $({\bf 9R})\,\,r=[cov(z_1,z_2)/\sqrt{D(z_1)D(z_2)}]$
leading to
\bq\label{9.16}
D(z_1)D(z_2)\geq\frac{(M(C))^2K^2}{4};\,\,K=\frac{1}{\sqrt{1-r^2}}
\end{equation}
Here K is analogous to the Schmidt number (cf. \cite{b12}, papers 6 and 7 and \cite{e2,l2,p1}
for details).  When $z_1=x$ and 
$z_2=p$,  C is unitary, and one obtains $({\bf 9S})\,\,D(x)D(p)\geq (K^2/4)$ leading to $\gD x\gD p\geq
(K/2)$.  We note that since $\hat{x}$ and $\hat{p}$ do not commute their quantum
covariance can not be estimated by their sampling as for a classical covariance; for
the corresponding estimate one needs the wave function $\psi(x)=\sqrt{\rho(x)}
exp(iS(x))$ which leads then to
\bq\label{9.17}
cov(x,p)=\frac{1}{2}<\psi|xp+px|\psi>=\int x\frac{\pp S(x)}{\pp x}\rho(x)dx
\end{equation}
where $\pp_xS=p$ is the momentum.  Fisher information and the Cramer-Rao inequality
are also discussed in \cite{b12}.
In summarizing one postulates in \cite{b12}
\begin{enumerate}
\item
The principle object of quantum informatics is a quantum system.  The evolution of the
quantum system is described via probability amplitudes which construct state vectors
in Hilbert space.
\item
The state vectors can be defined in different equivalent representations and are thus
connected by unitary transformations which describe the time evolution of the quantum
system.
\item
Measurements made in different unitary interconnected basis representations generate a set of mutually complementary statistical distributions.  For a fixed basis the square
of the absolute value of the probability amplitude defines the probability of the quantum
system detection in a corresponding basis state.
\item
The space for a composite system state is produced by the tensor product of the states
of individual systems.
\end{enumerate}
\indent
A conclusion reached here is that QM is a root statistical model, based not on the
actual probabilities but on their square roots; this is connected to the flow of 
half-densities under Bohmian flow in phase space as in \cite{g1,g14,g12,h4}.

\end{document}